\documentclass[preprint]{aastex}

\newcommand{\kms}{km~s$^{-1}$}

\newcommand{\subsun}{\mbox{$_{\odot}$}}
\newcommand{\teff}{$T_{eff}$}
\newcommand{\etal}{{\it et al.\/}}
\newcommand{\ew}{$W_{\lambda}$}

\begin{document}

\title{Pal 12 As A Part of the Sgr Stream; the Evidence
From Abundance Ratios\altaffilmark{1}}

\author{Judith G. Cohen\altaffilmark{2}}

\altaffiltext{1}{Based in part on observations obtained at the
W.M. Keck Observatory, which is operated jointly by the California 
Institute of Technology, the University of California, and the
National Aeronautics and Space Administration.}

\altaffiltext{2}{Palomar Observatory, Mail Stop 105-24,
California Institute of Technology, Pasadena, Ca., 91125, jlc@astro.caltech.edu}

\begin{abstract}
We present a detailed abundance analysis for 21 elements based on high
dispersion, high spectral resolution Keck spectra
for four members of the outer halo ``young'' Galactic globular cluster Pal 12.
All four stars show identical abundance distributions
with no credible indication of any star-to-star scatter.
However, the abundance ratios of the Pal 12 stars
are very peculiar.  There is no 
detected enhancement of the $\alpha$-elements;
the mean of [Si/Fe], [Ca/Fe] and [Ti/Fe] is $-0.07\pm0.05$ dex, O/Fe
is also Solar, while
Na is very deficient.  The distribution among
the heavy elements shows anomalies as well.
These 
are inconsistent with those of almost all Galactic globular clusters
or of field stars in the Galaxy.  The
peculiarities shown by the Pal 12 stars are, however, in good general
agreement with the trends established by
Smecker-Hane \& McWilliam and by Bonifacio \etal\ 
for stars in the Sgr dSph galaxy evaluated at the
[Fe/H] of Pal 12.  This reinforces
earlier suggestions that Pal 12 originally
was a cluster in the Sgr dSph galaxy which during the process of
accretion of this  galaxy by our own
was tidally stripped from the Sgr galaxy to become part of the 
extended Sgr stream.

\end{abstract}


\section{Introduction}

Pal 12 is a sparse globular cluster
(henceforth GC) located in the outer halo of
the Milky Way.  The CMD studies of \cite{gratton88}
and of \cite{stetson89} suggested that Pal 12 is probably
somewhat younger than the vast majority of Galactic GCs.
However, until there was a measurement of 
the metallicity of this GC, its age could not be 
determined robustly due to degeneracies in the CMD between age and
metallicity.  \cite{dacosta91} provided the crucial
datum; they found that
that Pal 12, in spite of its large galactocentric
distance, is quite metal rich, obtaining [Fe/H] = 
$-0.6$ dex\footnote{The 
standard nomenclature is adopted; the abundance of
element $X$ is given by $\epsilon(X) = N(X)/N(H)$ on a scale where
$N(H) = 10^{12}$ H atoms.  Then
[X/H] = log$_{10}$[N(X)/N(H)] $-$ log$_{10}$[N(X)/N(H)]\subsun, and similarly
for [X/Fe].}
from low resolution spectroscopy.  More recently,
\cite{brown97} carried out a high resolution
spectroscopic study of two stars in Pal 12, which yielded
[Fe/H] = $-1.0$ dex.

The combination of deep photometry and an estimate
of [Fe/H] led to the verification that Pal 12 is indeed
a young cluster.  
\cite{rosenberg98}, as part of their recent study of the 
age dispersion within the Galactic GC system \citep{rosenberg99},
suggest an age for Pal 12 of roughly 70\% that of the majority
of the halo GCs.  Assuming the latter group to be 12 Gyr old, 
they then infer an age of 8.4 Gyr for Pal 12, consistent with that of
the earlier CMD studies of \cite{gratton88} and \cite{stetson89}.

After the discovery of the Sgr dSph galaxy by \cite{ibata94},
several groups, including \cite{layden00}, noted
that four Galactic GCs (M54, Arp 2, Terzan 7 and Terzan 8), 
based on their positions on the sky,
appear to form a stream extending from the Sgr dSph, and suggested that these
GCs had been tidally stripped away from the Sgr galaxy.  
The GC M54 was postulated to be the
original nucleus of the Sgr galaxy.

\cite{irwin99} was the first to suggest that the GC Pal 12 
had also been tidally captured from the Sgr dSph galaxy by our Galaxy.
\cite{dinescu00} measured the proper motion
of this GC and calculated its orbit to find that
Pal 12's tidal capture from the Sgr dSph took place about
1.7 Gyr ago.  \cite{ibata01} demonstrated the existence of an extended
tidal stream of debris from the Sgr dSph galaxy in the Galactic halo,
which has recently been detected in the 2MASS  \citep{majewski03}
and in the SDSS \citep{ivezic03} databases.
Deep optical imaging over wide fields around Pal 12
by \cite{martinez02} and by \cite{bellazzini03} show that this GC
is embedded in the extended debris stream of stars torn
from the Sgr dSph.

The only existing high dispersion spectroscopic study of Pal 12
\citep*[that of][]{brown97}
was hitting the limits of what was observationally possible
with a 4-m telescope.  They analyzed the spectra of only the two brightest
probable members. Their primary result was that 
Pal 12 did not appear to show the enhancement
of the $\alpha$-process elements seen in
almost all GC stars.   Although this
is extremely interesting, the accuracy of their
analysis was limited by the quality of their spectra and some key
elements were not included.  The purpose of the present
paper is to provide a firmer foundation for their results, to
extend them as possible, and to compare the properties of
the Pal 12 stars with the recently published
abundance distributions for stars in the Sgr dSph galaxy of 
\cite{bonifacio00}, \cite{smecker02},
\cite{mcwilliam03} and of \cite{bonifacio03}.

\section{Stellar Sample and Stellar Parameters}

Since Pal 12 is potentially a young cluster in the outer halo,
there have been, as noted above, several recent CMD studies of Pal 12.
This is a rather sparse cluster, and there are  
only 4 probable members on the RGB brighter than $V \sim 16$ mag.
These are the four stars we have observed;
they are the same four
stars that were observed  at low resolution in the region
of the IR Ca triplet by by \cite{dacosta91}.
The stellar identifications we adopt are those of \cite{harris80}.

As in our earlier papers, we use the V-J and V-K colors
to establish the \teff\ for these stars.
We utilize the grid of predicted broad band colors and
bolometric corrections of \citet{houdashelt00}
based on the MARCS stellar atmosphere code of \citet{gus75}.  
In \cite{cohen01} we demonstrated that the Kurucz and MARCS 
predicted colors are essentially
identical, at least for the specific colors used here.
The optical photometry we adopt is from \cite{stetson89} and
the infrared photometry
is from 2MASS \citep{2mass}.  The reddening is low;
we adopt E(B-V) = 0.02 mag and a distance of 19.1 kpc
from the on-line database of
\cite{harris96}; the all-sky maps of \cite{schlegel98} 
yield a slightly larger E(B-V) of 0.036 mag.
It is important to note that the distance
was obtained with a knowledge of the probable young age of
this anomalous GC.

The surface gravities for the Pal 12 red giants
are calculated from their observed V
magnitudes, their \teff,
the cluster distance and the reddening, as in our earlier papers.  Here,
however, based on the isochrones of \cite{yi01},
we adopt a mass for the RGB stars of 1.0 M\subsun\
rather than the 0.8 M\subsun\ used in our earlier GC abundance
analyses; the latter
is appropriate for metal poor clusters with an age of 12 Gyr,
but not for Pal 12.

The most luminous Pal 12 giant, star S1, is somewhat cooler than
the minimum \teff\ in
the grid of \cite{houdashelt00}, and extrapolation beyond
the limit of this color grid was required.
The nominal \teff\ of 3850 K so inferred
from its observed colors gave poor ionization
equilibrium.  The photometric errors for our data produce a $\pm$50~K uncertainty
in \teff, hence we adopt a \teff\ for this star of 3900 K.
The resulting stellar parameters are listed in Table~\ref{table_teff}.
The \teff\ for the two stars analyzed by \cite{brown97} (the
cooler of the four stars analyzed here) are $\sim$60 K higher than those
adopted here; they use $V-I$ photometry in one case and
a $K$ mag for Star S1 from \cite{cohen78}; the older photometry
they use surely has uncertainties at least as large as that we use.  Given that,  
the agreement in \teff\ for the two stars in common seems reasonable.

\section{Observations}

All spectra were obtained with HIRES \citep{vogt94} at 
the Keck Observatory.  The four Pal 12 stars are too far
apart on the sky to fit two within the allowed 
HIRES slit length, and hence each had to be
observed individually.  The HIRES configuration used a 1.1 arcsec
wide slit (spectral resolution 34,000). Spectral coverage  extended
from 4650 to 7010~\AA, with small gaps in coverage between the echelle 
orders due to the current undersized HIRES detector.

The spectra were exposed to a SNR exceeding 100 per 4 pixel resolution
element in the continuum at the center of order 64 (about
5670~\AA).  This was calculated assuming Poisson statistics and ignoring
issues of cosmic ray removal, flattening etc.
These spectra were reduced using a combination 
of Figaro scripts \citep{shortridge93} and
the software package MAKEE \footnote{MAKEE was developed
by T.A. Barlow specifically for reduction of Keck HIRES data.  It is
freely available on the world wide web at the
Keck Observatory home page, http://www2.keck.hawaii.edu:3636/.}.
Details of the exposures are given in Table~\ref{table_sample}.
Heliocentric radial velocities were measured as described in
\cite{ramirez03}; all the stars are confirmed as members of Pal 12.
The four stars have a mean 
heliocentric $v_r$ of +28.9 \kms, with $\sigma$ = 0.8 \kms,
consistent with the presumed low mass of this cluster.  (The observational
error has not been removed from this observed velocity dispersion.)

The search for absorption features present in our HIRES data and the
measurement of their equivalent width (\ew) was done automatically with
a FORTRAN code, EWDET, developed for our globular cluster project. 
Details of this code and its features are described in \citet{ramirez01}.
Since we are observing only the most luminous (i.e. the coolest)
stars in a high metallicity cluster, considerable hand checking
of the equivalent widths had to be done at various stages of the analysis.

A list of unblended atomic lines 
with atomic parameters was created by 
merging our existing globular cluster list, developed from our
earlier work on M71 and M5, adding in bluer lines in part
from our work on very metal poor stars and in part as required
to fill in the bluer orders covered here.  We made extensive
use of the NIST Atomic Spectra 
Database Version 2.0 (NIST Standard Reference Database \#78,
see \citep{wei69,mar88,fuh88,wei96}).  The
online Solar spectrum taken with 
the FTS at the National Solar Observatory of \cite{wallace98}
and the set of Solar line identifications of 
\citet{moore66} were also used.
The list of lines identified and measured by EWDET is then correlated,
taking the radial velocity into account, 
to the list of suitable unblended lines  
to specifically identify the various atomic lines. 

All lines with \ew\ exceeding 200 m\AA\ were rejected, except
for the 6141.7 and 6496.9~\AA\ lines of Ba~II.  The even stronger
4934~\AA\ line of Ba~II was eliminated; it is blended and its 
\ew\ exceeded the cutoff in all the sample stars.
The list of equivalent widths used in the this analysis for
each of the four stars in Pal 12 is given in Table~\ref{table_eqw}.

To the maximum extent possible,
the  the atomic data and the analysis procedures
used here are identical to those developed in our earlier papers on M71
and on M5 (Cohen \etal\ 2001, Ram\'{\i}rez \etal\ 2001;
Ram\'{\i}rez \& Cohen 2002; Ram\'{\i}rez \& Cohen 2003).
For ions with hyperfine structure, we synthesize the spectrum for each line
including the appropriate HFS and isotopic components.
We use the HFS components from \cite{prochaska00} for the lines we utilize 
here of
Sc~II, V~I, Mn~I, Co~I.  For Ba~II, we adopt the HFS from \cite{mcwilliam98}.
We use the laboratory spectroscopy of \cite{lawler01a}
and \cite{lawler01b} to calculate the HFS patterns
for La~II and for Eu~II.  
We have updated our Nd~II
$gf$ values to those of \cite{denhartog03}.

We use the Solar abundances of \cite{grevesse98},
modified for the special cases of La~II, Nd~II and Eu~II
to those found by the respective recent laboratory studies
cited above.  In particular, this means we have adopted
$\epsilon$(Fe) for the Sun of 7.52 dex, although in our earlier papers
on M71 and M5 we used
7.44 dex for $\epsilon$(Fe)(Sun).

The microturbulent velocity ($v_t$) of a star can be determined 
spectroscopically by requiring the abundance to be independent of the 
strength of the lines.
We apply this technique here to the large sample of detected
Fe~I lines in each star, and use $v_t = 1.7$ or 1.8 \kms\ for the
four Pal 12 stars.

\section{Abundance Results for Pal 12}

Given the derived stellar parameters from Table~\ref{table_teff}, we 
determined the abundances using the equivalent widths obtained as 
described above.
The abundance analysis is carried out using a current version of the LTE
spectral synthesis program MOOG \citep{sneden73}.
We employ the grid of stellar atmospheres from \citet{kur93a} with
a metallicity of [Fe/H] = $-1.0$ dex\footnote{We use the grid of models without
convective overshoot.} to compute
the abundances of O, Na, Mg, Si, Ca, Sc, Ti, V, Cr, Mn, Fe, Co, Ni,
Cu, Zn, Y, Zr, Ba, La, Nd and Eu using the four stellar atmosphere
models with the closest \teff\ and log($g$) to each star's parameters.
The abundances were interpolated using results from the closest stellar model
atmospheres to the appropriate \teff\ and log($g$) for each star given
in \ref{table_teff}.  The results for the abundances of
these species in the four stars in Pal 12 are given in
Table~\ref{table_abund}.

Table~6 of \cite{ramirez02} (which discusses
M71, a GC of similar overall abundance) is a sensitivity table 
presenting the changes in deduced abundances of various species for small
changes in \teff, log($g$), and $v_t$; the entries in this table
for the cooler stars in M71 (\teff\ = 4250 K) can be used for
the Pal 12 stars as well. 

The ionization equilibrium for both Fe~I versus Fe~II and for
Ti~I versus Ti~II is satisfactory. 
The average difference for the four stars in Pal 12
between [Fe/H] as inferred from Fe~II lines
and from Fe~I lines is $+0.11\pm0.07$ dex and $+0.16\pm0.07$ dex for 
Ti\footnote{Due to the possible presence of random errors in stellar parameters,
all abundance ratios are assigned a minimum uncertainty of 0.05 dex.}.
The Fe ionization
equilibrium shifts by 0.2 dex for a 100 K change in \teff\ in this
temperature regime, hence a
systematic increase of the adopted \teff\ values
by 50 K would eliminate these small
discrepancies.  This possible systematic offset, for which no correction
has been made, is slightly smaller than the uncertainty in \teff.
It might result
from adopting a reddening for Pal 12 which is slightly too small.

Following upon our previous work, no non-LTE corrections have been applied
for the specific ions studied in the Pal 12 stars.  The detailed non-LTE
calculations of \cite{gratton99} and of \cite{takeda03} for the two Na~I doublets
we use suggest that for this regime of \teff\ the
non-LTE correction is about $+0.15$ dex.  For Ba~II, the 
non-LTE calculations of \cite{mashonkina99} and of
\cite{mashonkina00} suggest that a non-LTE correction
of $-$0.15 dex is appropriate for the metallicity of Pal 12
and the set of Ba~II lines we used.  In comparing with other
abundance analyses, the issue of implementing non-LTE
corrections and their adopted magnitudes must be considered.

\subsection{Comments on Individual Elements \label{sec_individual}}

The oxygen abundance is derived from the forbidden lines
at 6300 and 6363~\AA.  The subtraction of the night sky emission
lines at these wavelengths was reasonably straightforward. 
The stellar [O~I] lines in the spectra of the Pal 12 stars
are strong enough and the
radial
velocity of Pal 12 is sufficiently different from 0 \kms\ (and, for
the faintest Pal 12 star observed,
Pal 12 star 1305, the heliocentric correction provides additional
help in separating the separating the stellar and atmospheric [O~I] lines)
that their \ew\ can be reliably measured.
The C/O ratio was assumed to be Solar.
The O abundance is given with respect to [Fe/H] deduced
from lines
of Fe~I; the mean [O/Fe] becomes 0.09 dex smaller if expressed using the 
Fe~II lines instead.
The IR triplet at 7770~\AA\ is beyond the  wavelength range
of these spectra.    
No corrections were made for the Ni~I blend in the 6300~\AA\ 
discussed by \cite{allende01}; when this and the small difference
in adopted $gf$ value are taken into account, their value for the
Solar O abundance agrees with that adopted here to within 0.05 dex.

No lines of Al can be reached with this HIRES configuration;
the 6697~\AA\ double unfortunately falls in an interoder gap.

There are two detected lines of Cu, both of which have very large
HFS corrections, ranging between $-$0.5 and
$-1.0$ dex.  The abundance of copper in Pal 12 is therefore quite uncertain.

Three lines of Y~II are used here.  They are all crowded,
and the line at 5205.4~\AA\ is
too blended to use in all but the hottest star.  The $gf$ values are from
\cite{hannaford82}, and give good results for the Sun. In the hottest
star, the three lines gives reasonably consistent results.  However,
for star S1, the coolest star, the 4883.7~\AA\ line gives an abundance
more than a factor of 10 higher than the 5087.4~\AA\ line.  We are not
using any HFS corrections for Y.  \cite{hannaford82} suggest they
are small, but perhaps the  Y lines are so strong in this star that
use of HFS is required.  In computing
the mean Y abundance for Pal 12, we ignore star S1.

\subsection{Abundance Spreads}

We calculate the mean abundance for each atomic species ($X$) with observed
absorption lines for the four stars in Pal 12 as well as the 1$\sigma$ rms value
about that mean.  These values are given in the first 3 columns
of Table~\ref{table_abundsig}.  This is compared to the observational
uncertainty [$\sigma(obs)$], given in the fourth column of the Table.
$\sigma(obs)$ is taken as the uncertainty of the mean abundance
for a single star, i.e. the  1$\sigma$ rms value about the mean
abundance of species $X$ in a star/$\sqrt{N}$, where $N$ is the number of
observed lines of species $X$.  Values of $N$ can be found in
Table~\ref{table_abund}.
This definition of the observational uncertainty presumes that errors in the
\ew\ and the atomic data dominate; random errors in \teff\ or log($g$) are not
included.
Some species, an example being Fe~I
with its very large value of $N$, 
have very small values of $\sigma(obs)$;
we adopt a minimum of 0.05 dex for this parameter.

The ratio of these two different $\sigma$ values
is an indication of whether
there is any intrinsic star-to-star variation in [X/Fe].  A high value
of this ``spread ratio'' ($SR$), tabulated in the next to last column of
this table, suggests a high probability of intrinsic scatter for
the abundance of the species $X$.

The spectrum of the coolest Pal 12 star (S1) is the most crowded and blended
and also has many lines whose $W_{\lambda}$ exceeds the
cutoff  value of 200 m\AA.
For three species, this reduces the number of detected lines to only a single line
in this star, while the other three Pal 12 stars 
in our sample have two or more detected lines.
Thus in the case of Mg~I, Zn~I and Nd~II (as well as Y~II, see above), 
the averages and other statistics 
given in Table~\ref{table_abundsig} are calculated using only three stars, 
ignoring the coolest one.

Inspection of Table~\ref{table_abundsig} shows that for most species
$0.6 < SR \le 1.0$, indicating no sign of an intrinsic star-to-star range
in abundance.  Only
for  O, Cu, Zn, Zr, La and Nd does $SR$ exceed 1.0.  
Copper has by far the largest 
calculated $SR$ value; we believe that uncertainties associated with
the large HFS corrections are responsible for the large spread in the
derived Cu abundance among the four Pal 12 stars.
Ignoring Cu, all the remaining species 
listed above have only two detected lines at best (except
for Zr~I, which has 3), many of the detected rare earth lines
are weak, and some are crowded.  
We feel that when $N$ is very small ($N<4$), our estimate of $\sigma(obs)$
is biased low, and hence the $SR$ is biased high.  Thus
there is no credible evidence from our data for
an abundance spread for any species included in our analysis. 
In particular, the total range of 
[Mg/Fe] is only 0.02 dex (excluding star S1, where only 1 Mg~I line
could be used), the total range of [Si/Fe]
for all four Pal 12 stars is only 0.11 dex, etc.

Our derived [Fe/H] for Pal 12 is 0.2 dex higher than that
of \cite{brown97}, who found [Fe/H = $-1.0\pm0.1$ dex.  Only a small
part ($\sim0.05$ dex) of that difference can be attributed to the
cooler values of \teff\ adopted here. 
We add detections for the species O~I, V~I and Nd~II, not included
in the previous work on this GC, and
our abundances are much more precise than those of the
previous study. We have a sample large enough to assess the
possible presence of abundance spreads. 
To within their rather large errors,
the \cite{brown97} abundance ratios agree with those we have derived.
They correctly discerned the general nature of Pal 12's peculiarities.

\section{Comparison of the Abundances with Those of Other Stellar Systems}

In this section we compare the abundance distribution of Pal 12 with
those of various other galactic populations.  These patterns
reflect the history of star formation, the form of the initial mass function, and
the gas flows within  the Galaxy.  They provide signposts as to the 
nucleosynthesis in the early Universe and the nature of the
formation and collapse of the Galaxy.  Since the abundance distribution
of the stars in Pal 12 is so anomalous, we seek to find analogs of
it within other stellar components of the Galaxy.

\subsection{Comparison with Galactic GCs}

\begin{figure}
\epsscale{0.9}
\plotone{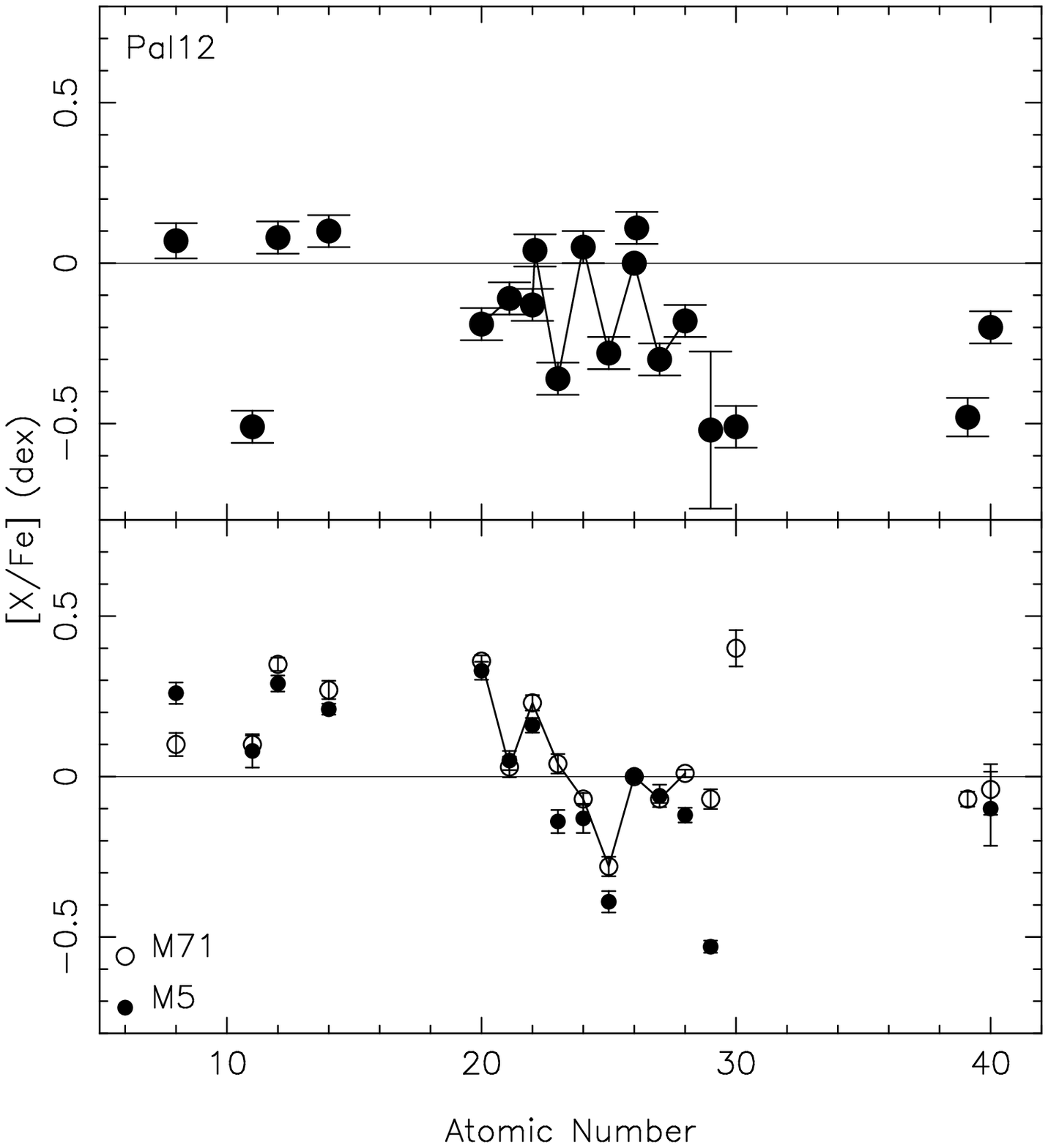}
\caption[]{The mean abundances of the four stars in Pal 12 are shown
as a function of atomic number from O through Zr in the upper panel.
The same information from large samples of stars in the GCs M71 and in M5
is shown  using
data from \cite{ramirez02} and from \cite{ramirez03}.  Those points representing
the abundances of species of consecutive atomic number from 20
to 28 for Pal 12 and for M71
are connected by line segments.
\label{figure_abund_low}}
\end{figure}

We compare the abundance ratios we have derived for Pal 12 with those
of typical Galactic GCs.  We adopt M71 and M5 as representative Galactic
GCs of suitable metallicity;
Pal 12 has [Fe/H] close to that of M71.
We use the data from \cite{ramirez02} for M71
and from \cite{ramirez03} for M5; these previous analyses 
from our group are almost directly comparable
with the present one for Pal 12 with no adjustments necessary. 
Figure~\ref{figure_abund_low} presents the abundance ratios as a function of atomic
number for species between oxygen and Zr.  Those of M71 and of M5
are shown in the lower panel.  

There are some features in common between the M71, M5 and
Pal 12 stars, particularly the large odd-even effect, which refers
to the unusually large deficiency of the odd atomic number species
(including  Na, Al, Sc, V, Mn, and Co) compared to their
even atomic number neighbors in the regime from Na to Ni in the periodic 
table\footnote{The Solar abundances themselves with respect to H, [X/H],
show a strong odd-even effect, but what we are referring to here is
an enhancement of this beyond what is characteristic of the Solar composition.}.
However, there are also real and important differences.
In particular, the $\alpha$ elements  Si, Ca, and Ti as well as O, 
are substantially
enhanced in M5 and in M71, as well as in essentially all GC stars, with
[$\alpha$/Fe] $\sim +0.3$ dex.  Pal 12,
on the other hand, shows no enhancement of these elements,
with the average of the mean values of [Si/Fe], [Ca/Fe] and
[Ti/Fe] being $-0.07\pm0.05$ dex;
[O/Fe] is also almost the Solar value in Pal 12 (see Table~\ref{table_abundsig}).
The deficiency of Na is very large in Pal 12
(the mean for the four Pal 12 stars of [Na/Fe] is $-0.51\pm0.04$ dex), 
much larger than is seen in
Galactic GCs.

\begin{figure}
\epsscale{0.9}
\plotone{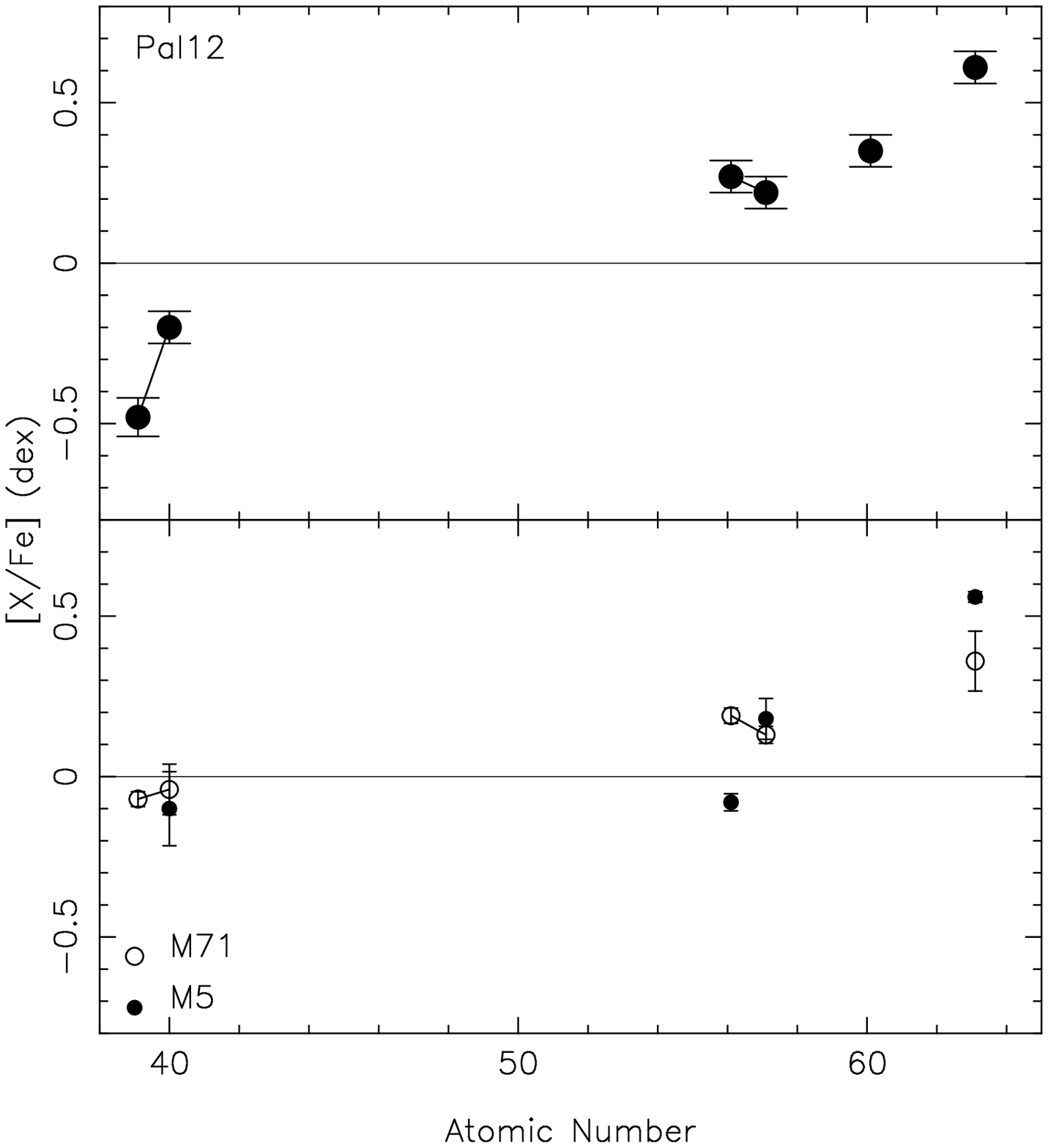}
\caption[]{The mean abundances of the four stars in Pal 12 are shown
as a function of atomic number from Y through Eu in the upper panel.
The same information from large samples of stars in the GCs M71 and in M5
is shown  using
data from \cite{ramirez02} and from \cite{ramirez03}.  Those points representing
the abundances of species of consecutive atomic number 
for Pal 12 and for M71 are connected by line segments.
\label{figure_abund_high}}
\end{figure}

Figure~\ref{figure_abund_high} shows the heavy elements from Y to Eu.
The presence of enhanced Eu in Pal 12 and in the comparison GCs M71 and M5
suggests a substantial contribution from the $r$-process.  The [Eu/Nd]
and [Eu/La] ratios  
ratios in Pal 12 are each within 0.2 dex of that of the Solar 
$r$-process ratio determined by \cite{burris00} from the
isotopic breakdowns of \cite{kappeler89}.
However, Y and Zr show substantially larger deficiencies
than are
seen in Galactic GCs.

\subsection{Comparison with Galactic Field Stars in the Halo and in the Disk}

We compare the properties of Pal 12 with those of Galactic disk and halo
stars.  \cite{reddy03} have completed an extremely accurate and internally
consistent analysis of 181 F and G   dwarfs the vast majority of which
belong to the Galactic thin disk.  They cover the metallicity range
$-0.7 \le [{\rm{Fe/H}}] \le +0.2$.
Their results are not consistent with the behavior of Pal 12,
which is only slightly
more metal poor than the metal-poor end of their sample.  
While the behavior of the $\alpha$-elements Si, Ca and Ti agree between the
disk star sample and Pal 12,  the behavior of Na is highly discrepant.
In the disk stars, [Na/Fe] rises from Solar to about +0.15 dex at the metal-poor
end of their sample, while that for Pal 12 is $-0.51\pm0.04$ dex.
The behavior of [Mn/Fe] is consistent between the disk stars and Pal 12, but
heavier than Fe, the
agreement deteriorates again.  The disk stars show [Zn/Fe] rising to reach about 
+0.1 dex
over this same range, while for Pal 12, [Zn/Fe] = $-0.51\pm0.13$ dex.
Y/Fe and Eu/Fe retain their Solar ratios throughout
the metallicity range covered by \cite{reddy03}, 
while those of Pal 12 are $-0.48\pm0.12$ and $+0.61\pm0.06$
respectively.  

A detailed study of a sample of stars from the thick disk has been
carried out by \cite{prochaska00}.  These stars, which reach to lower
metallicities than do the thin disk stars, also show large
enhancements of the $\alpha$ elements, with Na and Zn also enhanced.
Thus their abundance distribution also fails to match that of Pal 12.

\cite{fulbright02} and \cite{stephens99} have analyzed large samples of 
Galactic halo field stars.  \cite{fulbright02} 
attempted to correlate their kinematics  with their abundance distributions
He uses the Galactic rest frame velocities ($V_{RF}$), calculated from the $UVW$
velocities, removing the rotational velocity of the Local Standard of Rest,
to characterize the kinematic properties of the stars in his sample.
\cite{dinescu00} have measured the proper motion of Pal 12,
and suggest that the apogalactic radius for its orbit is $29.4\pm6.0$ kpc.
From their data we calculate  $V_{RF}(\rm{Pal~12})$ to be  $251\pm35$ \kms.   

\cite{fulbright02} finds that the highest
velocity stars in his sample have slightly lower $\alpha$-enhancements than
are typical of most halo stars.
A similar result was obtained by \cite{stephens02},
who concentrated on a sample of kinematically peculiar
stars in the outer halo.  These low $\alpha$-enhancements seen in a few 
stars which are probably in the outer Galactic halo
are not as low as those we observe in Pal 12.

The shift from the 
lighter to the heavier $s$-process elements between 
the first $s$-process (the Zr-peak) peak  towards the second peak at Ba
seen in Pal 12 is
also not matched by the halo stars.  The trends
for most element ratios with increasing ($V_{RF}$) found
among halo stars
by \cite{fulbright02} in general
have the right sign to reproduce the behavior of Pal 12 eventually, 
but fail to do so by significant amounts even for his highest velocity
bin, ($V_{RF} > 300$ \kms).

There are a very small number of metal-poor Galactic halo stars known
to have unusually low abundances of $\alpha$-elements, among the most extreme of
which is BD +80 245, which shows near-Solar $\alpha$/Fe with
[Fe/H $\sim -2$ dex \citep{carney97}.   However, even these stars,
a small heterogeneous group of which were studied by \cite{ivans03},
fail to match the abundance ratios seen among the Pal 12 stars.

\subsection{Comparison with the Sgr dSph Galaxy}

As shown above, the abundance distribution within Pal 12 fails
to match that of typical Galactic GCs, of most Galactic halo stars,
and of Galactic disk stars.  We next see how well it
matches that of stars within the Sgr dSph galaxy.  We utilize
the results of \cite{bonifacio00}, who analyzed two stars, and
of \cite{smecker02}, who analyzed 14 Sgr stars.  Results for [Mn/Fe]
from the latter sample are reported by \cite{mcwilliam03}. 
These analyses of the Sgr stars are similar to ours in that they
are classical LTE analyses and no corrections for non-LTE effects
were made.
\cite{bonifacio00}
characterizes their results for two stars with [Fe/H] $\sim0.2$ dex
by stating ``the abundance ratios found are
essentially Solar with a few exceptions: Na 
shows a strong overdeficiency, the heavy elements
Ba to Eu are overabundant, while Y is underabundant.''
This is a concise approximate description of the abundance ratios
in Pal 12.

\begin{figure}
\epsscale{0.9}
\plotone{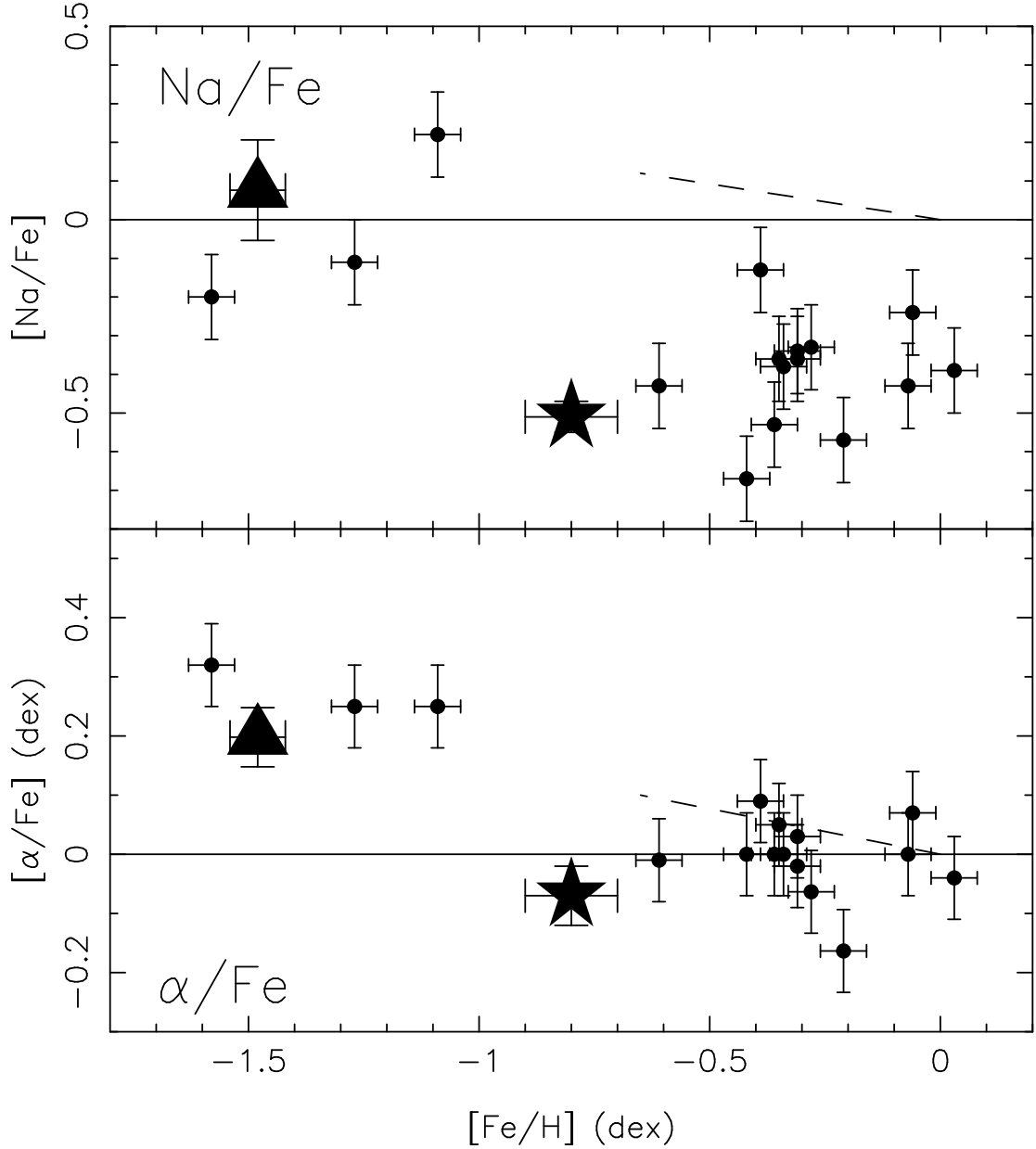}
\caption[]{The mean abundance of the $\alpha$-elements Si, Ca and Ti
with respect to Fe are shown as a function of [Fe/H] for a sample
of stars in the Sgr dSph galaxy with data from \cite{bonifacio00}
and from \cite{smecker02}.  Our result for Pal 12 is indicated by
the large star; that of the GC M54 is taken from \cite{brown99}.
The dashed line represents the behavior of the thin disk stars from
\cite{reddy03}.
The upper panel shows the same for [Na/Fe].
\label{figure_sgr_low}}
\end{figure}

We compare the mean $\alpha$ element ratio, [$\alpha$/Fe],
defined as the average of [Si/Fe],
[Ca/Fe] and [Ti/Fe], for stars in the Sgr dSph galaxy and for Pal 12.  
This is shown in the lower panel of
Figure~\ref{figure_sgr_low} as a function of metallicity.  The
Sgr system at low metallicity behave like typical Galactic GCs, while
at high metallicity, the $\alpha$-enhancement drops to zero.
The results of \cite{bonifacio03}, who analyze O, Mg, Si, Ca and Fe,
but not Ti,
in 10 additional Sgr stars covering the range 
$-0.85 <$ [Fe/H] $< +0.1$ dex, 
are similar.
Pal 12 fits right on the trend defined by the Sgr stars when the comparison
is made at the metallicity of Pal 12.
The upper panel shows the same for
[Na/Fe], where Pal 12 shares the tendency shown by the more metal-rich
Sgr stars to show very large depeletions of Na.
Also shown on this
figure is are the mean abundance ratios for a sample of five RGB stars in
M54, a Galactic GC 
long believed to have been tidally stripped from the Sgr dSph galaxy,
with data from \cite{brown99}.

The unusually low $\alpha$-ratios seen among the more metal-rich
of the Sgr dSph stars are also seen in other dwarf spheroidal galaxies.
Initial spectroscopic
analyses from Keck/HIRES spectra for abundances in the Draco and the
U Minor dSph galaxies have been carried out by \cite{shetrone98}
and by \cite{shetrone01}, with a total sample
of six stars in each of
these two galaxies.  In both of these galaxies,
the $\alpha$ elements (Si, Ca, Ti) appear to be less
enhanced relative to Fe than they are in the Galactic halo field.
The four southern dSph galaxies with small samples
of stars studied with UVES  
\citep{shetrone03,tolstoy03} (Sculptor, Fornax, Carina and Leo I)
also show this pattern.
Thus this is now a well established result for the metal-rich component
of the dSph satellites of the Galaxy.

\cite{nissen97} suggested that the large depletion
of Na seen in some $\alpha$-poor halo stars
is accompanied by a smaller depletion of Ni.
\cite{nissen04} notes that this correlation extends  to include
stars in the Sgr and other local dSph galaxies, and again Pal
12 fits right on the trend at its metallicity.  The possible
origin of this odd coupling of a
presumed  $\alpha$-element with a presumed Fe-peak element
is discussed in their papers.

\begin{figure}
\epsscale{0.7}
\plotone{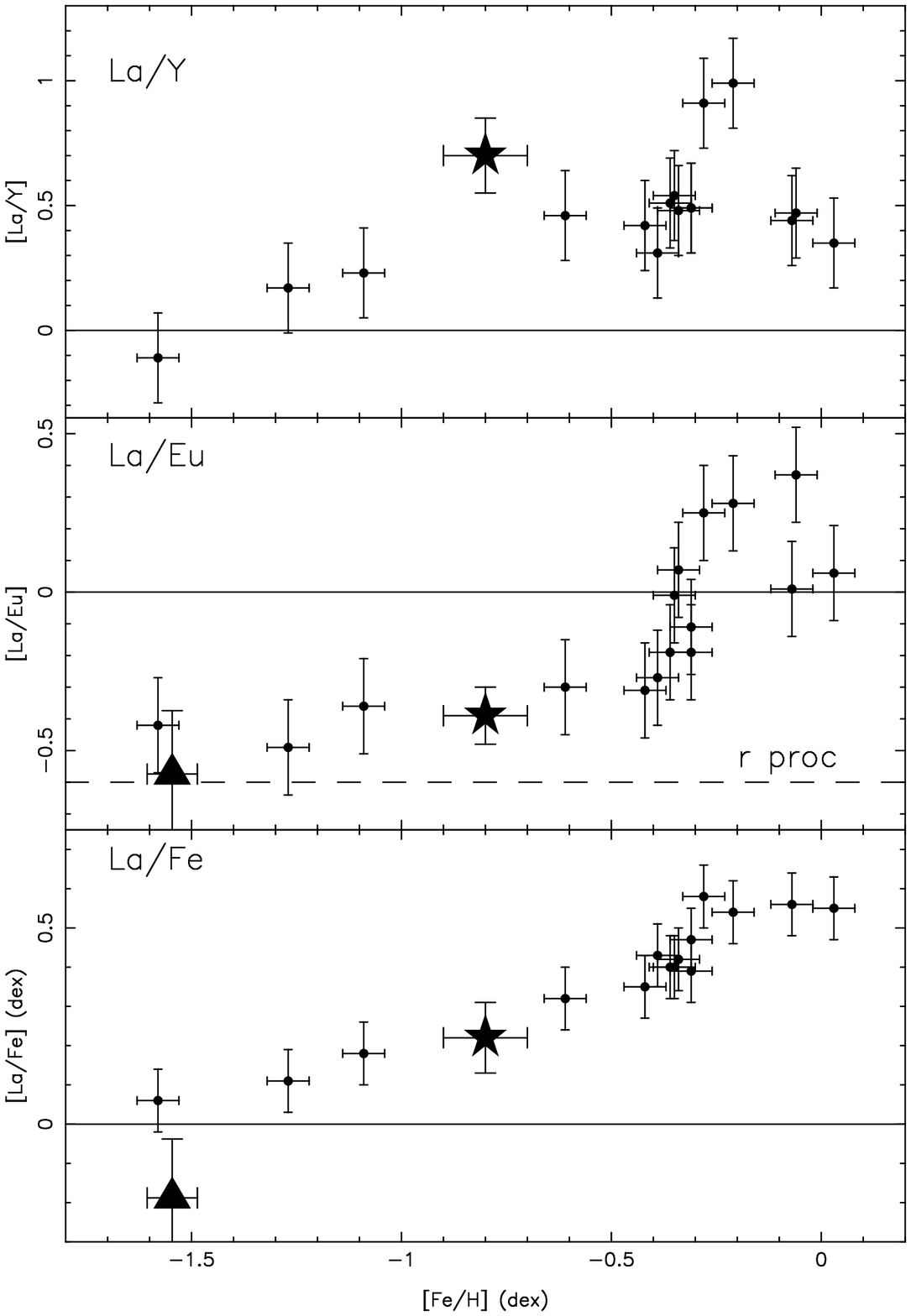}
\caption[]{The abundance ratio [La/Fe] 
is shown as a function of [Fe/H] for a sample
of stars in the Sgr dSph galaxy with data from \cite{bonifacio00}
and from \cite{smecker02} in the bottom panel.  Our result for Pal 12 is indicated by
the large star; that of the GC M54 is taken from \cite{brown99}.
The middle panel shows the same for [La/Eu], while the
top panel displays [La/Y].  The dashed line in the middle panel
indicates the [La/Eu] ratio from the Solar $r$-process.
\label{figure_abund_sgr_high}}
\end{figure}

Figure~\ref{figure_abund_sgr_high} shows the ratios 
[La/Fe], [La/Eu] and [La/Y] as a function of [Fe/H]
in a manner similar to that of Figure~\ref{figure_sgr_low}.  
Again the Sgr stars show a systematic trend of each ratio with metallicity.
In each panel of this figure,
both the Pal 12 point and the M54 point lie on the trend 
at the  metallicity appropriate for the GC.  The shift of the La/Eu
ratio between low and high metallicity Sgr stars (with Pal 12 behaving
like a low metallicity Sgr star) is equivalent to the shift between
$r$ and $s$-process dominance among Galactic halo stars  
\citep*[see, e.g.][]{burris00}.

Among the set of four GCs long suspected to have been
stripped from the Sgr dSph galaxy, Arp 2, like M54, is metal poor.  
At low metallicity, both the Sgr stars and M54 show abundance 
ratios similar to typical Galactic GCs. It is only among the more metal
rich Sgr stars that differences emerge, which are shared by
Pal 12.  Terzan 7 is also suspected to be metal rich, but a preliminary
study of 3 luminous RGB stars in this cluster 
by \cite{wallerstein02} claimed to see the $\alpha$-enhancements
typical of GCs.  However, reanalysis of this material by 
\cite{sbordone03} and by \cite{tau03} gives
[Fe/H] = $-0.6$ dex, with Solar $\alpha$/Fe,
consistent with our Pal 12 result and with the run of the Sgr dSph stars.

\subsection{The Age-Metallicity Relation for Sgr}

Here we review the age-metallicity relation for the Sgr dSph galaxy
and demonstrate that the measured metallicity and inferred age 
for Pal 12 are consistent with that, to within the large uncertainties.
Ages for GCs can be determined through isochrone fitting or through
differences between the HB and the main sequence turnoff from
suitable multi-color photometry.  The age of Pal 12 has been determined by
\cite{gratton88}, \cite{stetson89}, and most recently
by \cite{rosenberg98}.  Ages for the four GCs which have long been 
believed to be
associated with the Sgr galaxy (M54, Arp 2, Terzan 7 and Terzan 8)
are given by \cite{layden00}, as are the most recent abundance determinations
for these objects.  \cite{smecker02} have estimated ages for the 14 stars
in their Sgr sample; these 
determinations for individual stars
have much larger uncertainties than those for the GCs
and are further compromised by the possible depth of the galaxy
along the line of sight.

\begin{figure}
\epsscale{0.9}
\plotone{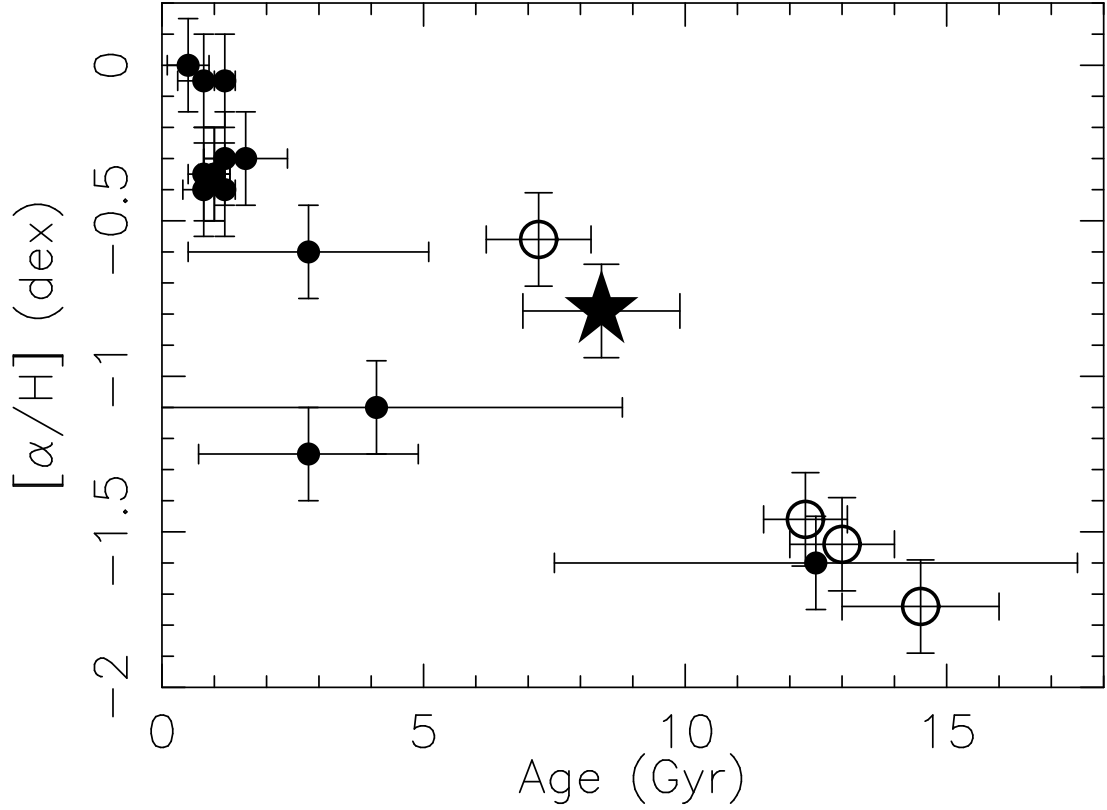}
\caption[]{[$\alpha$/H] versus age is shown for  stars in the Sgr dSph galaxy
\citep*[data from][]{smecker02} (small filled circles), the four Galactic GCs believed to
be associated with the Sgr stream \citep*[data large from][]{layden00}
(large open circles), and for Pal 12 (large star).
\label{figure_abund_age}}
\end{figure}

Figure~\ref{figure_abund_age} illustrates the age-metallicity (in the
form of [$\alpha$/H]) relation for the Sgr dSph stars with abundance
analyses by  \cite{smecker02} and for these four Galactic GCs,
as well as for Pal 12.  While the ages for the Sgr stars are
very uncertain, the location of Pal 12 in this figure appears consistent
with the age-metallicity relationship displayed by the stars in the Sgr
dSph galaxy.  An analysis of more Sgr stars of varying ages and
metallicities is needed to better define the age-metallicity
relationship of the Sgr galaxy to refine this comparison.

\subsection{Comments on Nucleosynthesis}

The general principles of nucleosynthesis of the elements in stars
are reviewed by \cite{wheeler89} and by \cite{mcwilliam97}.
Some of the trends described above are fairly easy to
explain in this context.  
The high $\alpha$-element ratios seen in Galactic GCs and
halo stars are ascribed to a very old population for which
there was insufficient time for 
type I SN to have evolved and detonated and hence contributed their 
nuclear processed material (consisting mainly of Fe-peak
elements) to the ISM.  The relevant minimum timescale for this is
difficult to estimate since the evolution of close binary systems containing
a white dwarf, believed to be the progenitors of Type Ia SN, must be followed
in detail including mass accretion; a recent attempt
to carry out this calculation by \cite{han03} gives
a minimum timescale of $\sim3 \times 10^8$ yr.
The low $\alpha$-ratios seen in Pal 12 must thus be 
ascribed a region with an star formation ongoing for a few
Gyr, thus achieving full contributions from both Type I and Type II SN.
Na is largely synthesized in massive stars, even 
more massive than those producing the bulk of the
$\alpha$-elements, which subsequently explode
as Type II SN. The extremely large deficiency of Na seen in Pal 12 
thus is probably
consistent with the low $\alpha$-element ratios and may not require
an IMF with a deficit of very massive stars during the
initial epochs of star formation.

The odd-even effect (see Figure~\ref{figure_abund_low}) is prominent
in the region of the periodic table from Na to Co.
Elements with even atomic numbers in this region of 
the periodic table all
have their most abundant stable isotope containing even numbers of both
protons and neutrons.  
\cite{arnett71} \citep*[see also][]{arnett96} discussed the odd-even effect 
for production of these elements through explosive nucleosynthesis.
He demonstrated that the
amplitude of this effect depends on the neutron excess, with larger
amplitude for smaller neutron excesses which are characteristic
of metal-poor material that has not been modified by H or He burning.  
It has long
been clear that the odd-even effect is strongly enhanced among halo stars,
and this effect is seen in Pal 12 as well.

While the presence of enhanced Eu requires a substantial 
$r$-process contribution
in Pal 12, there may be contributions 
to its inventory of heavy elements from the $s$-process as well.  
The other key feature
for the heavy elements in Pal 12 is the excess
deficiency of Y and Zr compared to Ba and La.  
The $s$-process elements in stars with metallicities above $-1.5$ dex 
are believed to be formed mostly in AGB stars of intermediate mass.  
Nucleosynthesis in such conditions, reviewed by \cite{busso99}, with
more recent calculations by \cite{busso01},
is dependent on the ratio of free nuetrons to seed nuclei, presumably Fe.
A larger neutron-to-seed ratio will lead to increased production of the heavier
$s$-process elements relative to the lighter ones, i.e. to those
in the Ba peak versus those near Sr, which is what is seen among
the Pal 12 stars. In an extreme case, this leads to $s$-process production  
of detectable amounts of lead in very low metallicity stars 
\citep*[see, e.g.][]{cohen03}.
Thus the anomalous ratios seen among the heavy elements in Pal 12 
may be symptomatic
of $s$-process nucleosynthesis at low metallicities.

\cite{smecker02} give a more detailed discussion of these ideas and how
they apply to their sample of Sgr dSph stars spanning a wide range
of metallicity, and showing peculiarities which depend on their [Fe/H].

\section{Summary}

We present a detailed abundance analysis for 21 elements based on high
dispersion, high spectral resolution Keck/HIRES spectra
for four members of the outer halo ``young'' Galactic GC Pal 12
which has an age of $\sim$8 Gyr, $\sim30\%$ younger than almost all GCs
\citep{rosenberg98}, and is known to have a rather high metallicity for a GC
in the outer halo \citep{brown97}.  Since the discovery of the Sgr
dSph galaxy, the Galactic GCs
M54, Arp 2, Terzan 7 and Terzan 8
have been believed to be associated with the this  galaxy, which is
currently being accreted by the Milky Way.  \cite{irwin99} recently
suggested that Pal 12 is also a Galactic GC which has been tidally stripped
from the Sgr galaxy.

All four stars in our sample in Pal 12 show identical abundance distributions
with no credible indication of any star-to-star scatter.
However, the abundance ratios of the Pal 12 stars
are very peculiar.  With $[{\rm{Fe/H}}] \sim -0.8$ dex, there is no 
detected enhancement of the $\alpha$-elements;
the mean of [Si/Fe], [Ca/Fe] and [Ti/Fe] is $-0.07\pm0.05$ dex, O/Fe
is also Solar, while
Na is very deficient.  The distribution among
the heavy  elements shows anomalies as well, with a shift from the
first $s$-process peak (Y and Zr) to the second peak (Ba and La).
(Eu is highly enhanced, as is typical in Galactic GCs of similar
metallicity.)
We show that these abundance peculiarities are not seen among
other stellar populations in the Galaxy, including
almost all Galactic GCs
and field stars in the disk, halo and bulge of the Galaxy.  

The abundance anomalies shown by the Pal 12 stars are, however, in good general
agreement with the trends established by \cite{smecker02} and
\cite{bonifacio00}
for stars in the Sgr dSph galaxy, when evaluated at the
[Fe/H] of Pal 12.  It is interesting to note that the trend of
low/no $\alpha$-enhancement has been found in all high metallicity
stars in all the dSph satellites of the Galaxy studied to date.
The abundance peculiarities exhibited by Pal 12
can in general be explained by initiating star formation 
in a metal poor system, and having
an extended period of star formation lasting at least a few Gyr.

Our abundance analysis of a sample of four stars in Pal 12 thus
reinforces
earlier suggestions that this GC originally
was a cluster in the Sgr dSph galaxy which during the process of
accretion of this  galaxy by our own
was tidally stripped from the Sgr galaxy to become part of the 
extended Sgr stream.  
Further searches for such anomalies are perhaps best concentrated among
metal rich outer halo Galactic GCs; in the metal poor ones the
duration of the major epoch of star formation may have been too brief 
for significant differences to develop in
star formation history and hence in abundance distributions
among the GCs.

\acknowledgements
This paper is for Chris Sneden, who at his review talk in Feb. 2003 at
the Carnegie symposium on ``The Origin and Evolution of the Elements''
appealed for someone to take another look
at Pal 12.
The entire Keck/HIRES user communities owes a huge debt to 
Jerry Nelson, Gerry Smith, Steve Vogt, and many other 
people who have worked to make the Keck Telescope and HIRES  
a reality and to operate and maintain the Keck Observatory. 
We are grateful to the W. M.  Keck Foundation for the vision to fund
the construction of the W. M. Keck Observatory. 
The author wishes to extend special thanks to those of Hawaiian ancestry
on whose sacred mountain we are privileged to be guests. 
Without their generous hospitality, none of the observations presented
herein would have been possible.
We are grateful to the National Science Foundation for partial support under
grant AST-0205951 to JGC.  We thank Jason Prochaska and Andy McWilliam
for providing their tables of hyperfine structure in digital form, and Andy
McWilliam and Tammy Smecker-Hane for providing their Sgr dSph abundances
in digital form in advance of publication.

\clearpage



\clearpage

%
%
\clearpage
\begin{deluxetable}{lrrrrr}
\tablenum{1}
\tablewidth{0pt}
\tablecaption{The Sample of Stars in Pal 12. \label{table_sample}}
\tablehead{
\colhead{ID\tablenotemark{a}} & 
\colhead{V\tablenotemark{b}} &
\colhead{Date Obs.} & 
\colhead{Exp. Time} &
\colhead{SNR\tablenotemark{c}} &
\colhead{$v_{r}$} \\ 
\colhead{} & 
\colhead{(mag)} &   
\colhead{} & 
\colhead{(sec)} &
\colhead{} & \colhead{(\kms)} }
\startdata 
S1          & 14.57 & 6/2003 & 1500 & $>100$ & +28.8  \\
1118        & 14.84 & 6/2003 & 1800 & $>100$ & +30.0  \\
1128        & 15.43 & 6/2003 & 1500 & $>100$ & +28.1  \\
1305        & 15.86 & 8/2003 & 3600 & $>100$ & +28.8  \\
\enddata
\tablenotetext{a}{Identifications are from \cite{harris80}.}
\tablenotetext{b}{V photometry from \cite{stetson89}.}
\tablenotetext{c}{Signal to noise ratio in the continuum near
5865 \AA\ per 4 pixel spectral resolution element.}
\end{deluxetable}

%
%
\clearpage
\begin{deluxetable}{lccc}
\tablenum{2}
\tablewidth{0pt}
\tablecaption{Stellar Parameters for the Pal 12 Sample. 
\label{table_teff}}
\tablehead{
\colhead{ID\tablenotemark{a}} &
\colhead{\teff } &
\colhead{log($g$) } &
\colhead{$v_t$} \\
\colhead{} &
\colhead{(K)} &
\colhead{(dex)} &
\colhead{(km/s)} 
}
\startdata
S1   &  3900 &  0.63  &  1.8 \\
1118 &  4000 &  0.84  &  1.8 \\
1128 & 4260 & 1.30 & 1.7 \\
1305 &  4465 &  1.62  &  1.7 \\
\enddata
\tablenotetext{a}{Identifications are from \cite{harris80}.}
\end{deluxetable}

%
%
\clearpage                                                                      
\begin{deluxetable}{llrrcccc}                             
\tablenum{3}                                                                    
\tablewidth{0pt}                                                                
\tablecaption{Equivalent Widths For the Pal 12 Stars \label{table_eqw}}                      
\tablehead{                                                                     
\colhead{Ion} & \colhead{$\lambda$} & \colhead{$\chi$} & 
\colhead{log $gf$} & \colhead{$W_{\lambda}$(S1)} &
\colhead{$W_{\lambda}$(1118)} &  \colhead{$W_{\lambda}$(1128)} &
\colhead{$W_{\lambda}$(1305)} \\
\colhead{} & \colhead{($\AA$)} & \colhead{(eV)} & \colhead{} &
\colhead{(m$\AA$)} & \colhead{(m$\AA$)} & \colhead{(m$\AA$)} & 
\colhead{(m$\AA$)} 
}
\startdata
  OI   &   6300.30 &   0.00 &  $-$9.78 &   68 &   79 &   39 &   39   \\ 
  OI   &   6363.78 &   0.02 & $-$10.30 &   45 &   40 &   14 &   18   \\ 
  NaI  &   5682.63 &   2.10 &  $-$0.70 &  113 &  103 &   77 &   67   \\ 
  NaI  &   5688.19 &   2.10 &  $-$0.42 &  128 &  108 &  100 &   94   \\ 
  NaI  &   6154.23 &   2.10 &  $-$1.53 &   41 &   35 &   19 &   15   \\ 
  NaI  &   6160.75 &   2.00 &  $-$1.23 &   65 &   53 &   30 &   28   \\ 
  MgI  &   4703.00 &   4.34 &  $-$0.67 &  \nodata &  184 &  191 &  185   \\
  MgI  &   5528.40 &   4.34 &  $-$0.48 &  \nodata &  \nodata &  \nodata &  199   \\
  MgI  &   5711.09 &   4.34 &  $-$1.67 &  119 &  129 &  111 &  102   \\ 
  SiI  &   5665.55 &   4.92 &  $-$2.04 &   43 &   40 &   54 &   44   \\ 
  SiI  &   5690.43 &   4.93 &  $-$1.87 &   40 &   36 &   38 &   44   \\ 
  SiI  &   5701.10 &   4.93 &  $-$2.05 &   45 &   42 &   39 &   42   \\ 
  SiI  &   5772.15 &   5.08 &  $-$1.75 &   30 &   32 &   47 &   51   \\ 
  SiI  &   5793.07 &   4.93 &  $-$2.06 &   37 &   27 &   32 &   34   \\ 
  SiI  &   5948.54 &   5.08 &  $-$1.23 &   69 &   78 &   75 &   80   \\ 
  SiI  &   6145.02 &   5.61 &  $-$1.44 &   17 &   18 &   24 &   25   \\ 
  SiI  &   6155.13 &   5.62 &  $-$0.76 &   61 &   65 &   55 &   62   \\ 
  SiI  &   6237.32 &   5.62 &  $-$1.01 &   33 &   31 &   38 &   42   \\ 
  SiI  &   6721.84 &   5.86 &  $-$0.94 &  \nodata &   22 &   16 &   22   \\ 
  SiI  &   7003.57 &   5.96 &  $-$0.83 &   19 &   18 &   34 &   30   \\ 
  SiI  &   7005.89 &   5.98 &  $-$0.73 &   37 &   28 &   46 &   36   \\ 
  SiI  &   7034.90 &   5.87 &  $-$0.88 &   33 &   21 &   34 &   49   \\ 
  CaI  &   5512.99 &   2.93 &  $-$0.27 &  119 &  115 &   97 &   95   \\ 
  CaI  &   5581.96 &   2.52 &  $-$0.47 &  154 &  143 &  126 &  114   \\ 
  CaI  &   5588.75 &   2.52 &   0.44 &  \nodata &  193 &  174 &  163   \\ 
  CaI  &   5590.11 &   2.52 &  $-$0.71 &  154 &  143 &  120 &  111   \\ 
  CaI  &   5601.28 &   2.52 &  $-$0.44 &  177 &  163 &  140 &  125   \\ 
  CaI  &   6156.02 &   2.52 &  $-$2.19 &   47 &   42 &   30 &   14   \\ 
  CaI  &   6161.30 &   2.52 &  $-$1.03 &  154 &  136 &  108 &   90   \\ 
  CaI  &   6166.44 &   2.52 &  $-$1.05 &  142 &  121 &  107 &   88   \\ 
  CaI  &   6169.04 &   2.52 &  $-$0.54 &  158 &  143 &  124 &  111   \\ 
  CaI  &   6169.56 &   2.52 &  $-$0.27 &  175 &  162 &  142 &  127   \\ 
  CaI  &   6471.66 &   2.52 &  $-$0.59 &  155 &  140 &  126 &  119   \\ 
  CaI  &   6493.78 &   2.52 &   0.14 &  180 &  164 &  156 &  147   \\ 
  CaI  &   6499.65 &   2.54 &  $-$0.59 &  153 &  148 &  122 &  111   \\ 
  CaI  &   6508.85 &   2.52 &  $-$2.12 &   48 &   32 &   16 &   12   \\ 
  ScII &   5526.79 &   1.77 &   0.13 &  114 &  111 &  102 &  104   \\ 
  ScII &   5657.90 &   1.51 &  $-$0.50 &  113 &  120 &  106 &  103   \\ 
  ScII &   5667.15 &   1.50 &  $-$1.24 &   80 &   84 &   62 &   61   \\ 
  ScII &   5669.04 &   1.50 &  $-$1.12 &  \nodata &  \nodata &   81 &   87   \\ 
  ScII &   5684.20 &   1.51 &  $-$1.08 &   79 &   76 &   71 &   64   \\ 
  ScII &   6245.64 &   1.51 &  $-$1.13 &   86 &   78 &   70 &   74   \\ 
  ScII &   6604.60 &   1.36 &  $-$1.48 &   91 &   76 &   71 &   63   \\ 
  TiI  &   4681.92 &   0.85 &  $-$1.07 &  \nodata &  \nodata &  195 &  \nodata   \\ 
  TiI  &   4981.74 &   0.85 &   0.50 &  \nodata &  \nodata &  187 &  170   \\ 
  TiI  &   5022.87 &   0.83 &  $-$0.43 &  \nodata &  199 &  153 &  131   \\ 
  TiI  &   5039.96 &   0.02 &  $-$1.13 &  \nodata &  \nodata &  187 &  145   \\ 
  TiI  &   5426.26 &   0.02 &  $-$3.01 &  160 &  143 &   82 &   55   \\ 
  TiI  &   5471.20 &   1.44 &  $-$1.39 &   91 &   69 &   44 &   26   \\ 
  TiI  &   5474.21 &   1.46 &  $-$1.23 &  134 &   96 &   61 &   42   \\ 
  TiI  &   5490.15 &   1.46 &  $-$0.93 &  135 &  121 &   77 &   60   \\ 
  TiI  &   5648.57 &   2.49 &  $-$0.25 &   57 &   45 &   24 &   16   \\ 
  TiI  &   5662.16 &   2.32 &  $-$0.11 &  117 &  102 &   72 &   53   \\ 
  TiI  &   5689.49 &   2.30 &  $-$0.47 &   73 &   63 &   41 &   28   \\ 
  TiI  &   5702.69 &   2.29 &  $-$0.57 &   70 &   56 &   38 &   21   \\ 
  TiI  &   5739.46 &   2.25 &  $-$0.60 &   56 &   39 &   20 &   19   \\ 
  TiI  &   5739.98 &   2.24 &  $-$0.67 &   49 &   39 &   18 &   17   \\ 
  TiI  &   5866.45 &   1.07 &  $-$0.84 &  195 &  167 &  129 &   98   \\ 
  TiI  &   5880.27 &   1.05 &  $-$2.05 &  120 &  104 &   55 &   39   \\ 
  TiI  &   5922.11 &   1.05 &  $-$1.47 &  154 &  126 &   90 &   64   \\ 
  TiI  &   5937.81 &   1.07 &  $-$1.89 &  114 &   93 &   55 &   38   \\ 
  TiI  &   5941.75 &   1.05 &  $-$1.52 &  152 &  134 &   93 &   82   \\ 
  TiI  &   5953.16 &   1.89 &  $-$0.33 &  137 &  126 &   92 &   71   \\ 
  TiI  &   5965.83 &   1.88 &  $-$0.41 &  143 &  121 &   88 &   78   \\ 
  TiI  &   5978.54 &   1.87 &  $-$0.50 &  121 &  109 &   81 &   66   \\ 
  TiI  &   6064.63 &   1.05 &  $-$1.94 &  119 &   98 &   64 &   37   \\ 
  TiI  &   6091.17 &   2.27 &  $-$0.42 &   87 &   75 &   54 &   32   \\ 
  TiI  &   6092.80 &   1.07 &  $-$1.38 &   66 &   39 &   27 &  \nodata   \\ 
  TiI  &   6126.22 &   1.07 &  $-$1.42 &  158 &  135 &   98 &   74   \\ 
  TiI  &   6258.10 &   1.44 &  $-$0.35 &  \nodata &  177 &  126 &  103   \\ 
  TiI  &   6258.71 &   1.46 &  $-$0.24 &  \nodata &  \nodata &  186 &  134   \\ 
  TiI  &   6261.10 &   1.43 &  $-$0.48 &  \nodata &  \nodata &  144 &   98   \\ 
  TiI  &   6303.76 &   1.44 &  $-$1.57 &  107 &   92 &   54 &   28   \\ 
  TiI  &   6312.22 &   1.46 &  $-$1.55 &   96 &   82 &   48 &   34   \\ 
  TiI  &   6743.12 &   0.90 &  $-$1.63 &  168 &  144 &   99 &   73   \\ 
  TiII &   6861.45 &   1.24 &  $-$0.74 &   64 &   54 &   18 &  \nodata   \\ 
  TiII &   4657.20 &   1.24 &  $-$2.32 &  \nodata &  \nodata &  108 &  \nodata   \\ 
  TiII &   4708.67 &   1.24 &  $-$2.37 &  112 &  100 &  103 &   97   \\ 
  TiII &   4865.62 &   1.12 &  $-$2.81 &   88 &  103 &   91 &   91   \\ 
  TiII &   4911.20 &   3.12 &  $-$0.34 &   88 &   76 &   68 &   77   \\ 
  TiII &   5185.91 &   1.89 &  $-$1.46 &  \nodata &  110 &  113 &  107   \\ 
  TiII &   5336.79 &   1.58 &  $-$1.63 &  126 &  117 &  113 &  122   \\ 
  VI   &   5670.85 &   1.08 &  $-$0.43 &  156 &  137 &   94 &   55   \\ 
  VI   &   5703.57 &   1.05 &  $-$0.21 &  154 &  133 &   94 &   76   \\ 
  VI   &   6081.44 &   1.05 &  $-$0.58 &  141 &  119 &   82 &   46   \\ 
  VI   &   6090.22 &   1.08 &  $-$0.06 &  159 &  138 &  105 &   74   \\ 
  VI   &   6199.20 &   0.29 &  $-$1.28 &  \nodata &  175 &  108 &   59   \\ 
  VI   &   6243.10 &   0.30 &  $-$0.98 &  \nodata &  \nodata &  140 &   90   \\ 
  VI   &   6251.82 &   0.29 &  $-$1.34 &  185 &  158 &  106 &   65   \\ 
  VI   &   6274.64 &   0.27 &  $-$1.67 &  151 &  131 &   76 &   46   \\ 
  VI   &   6285.14 &   0.28 &  $-$1.51 &  158 &  132 &   84 &   56   \\ 
  VI   &   6504.16 &   1.18 &  $-$1.23 &   76 &   59 &   43 &   19   \\ 
  CrI  &   4652.17 &   1.00 &  $-$1.03 &  \nodata &  \nodata &  166 &  168   \\ 
  CrI  &   5345.81 &   1.00 &  $-$0.97 &  \nodata &  \nodata &  199 &  174   \\ 
  CrI  &   5348.33 &   1.00 &  $-$1.29 &  \nodata &  \nodata &  170 &  166   \\ 
  CrI  &   5783.09 &   3.32 &  $-$0.50 &  101 &   60 &   50 &   37   \\ 
  CrI  &   5783.89 &   3.32 &  $-$0.29 &  107 &   96 &   77 &   71   \\ 
  CrI  &   5785.02 &   3.32 &  $-$0.38 &  \nodata &  \nodata &   45 &   68   \\ 
  CrI  &   5787.96 &   3.32 &  $-$0.08 &   93 &   79 &   71 &   59   \\ 
  CrI  &   5844.59 &   3.01 &  $-$1.76 &   37 &   35 &   21 &   13   \\ 
  CrI  &   6978.49 &   3.46 &   0.14 &  121 &  115 &   91 &   71   \\ 
  CrI  &   6979.80 &   3.46 &  $-$0.41 &   75 &   64 &   50 &   39   \\ 
  MnI  &   4754.04 &   2.28 &  $-$0.09 &  199 &  192 &  163 &  151   \\ 
  MnI  &   4783.42 &   2.30 &   0.04 &  \nodata &  \nodata &  172 &  174   \\
  MnI  &   4823.50 &   2.32 &   0.14 &  \nodata &  \nodata &  \nodata &  150   \\
  MnI  &   5537.74 &   2.19 &  $-$2.02 &  \nodata &  \nodata &   92 &   54   \\ 
  MnI  &   6021.80 &   3.08 &   0.03 &  156 &  149 &  127 &  114   \\ 
  FeI  &   4788.77 &   3.24 &  $-$1.81 &  \nodata &  \nodata &   92 &   90   \\ 
  FeI  &   5083.34 &   0.96 &  $-$2.96 &  \nodata &  \nodata &  182 &  183   \\ 
  FeI  &   5198.72 &   2.22 &  $-$2.14 &  \nodata &  196 &  158 &  154   \\ 
  FeI  &   5393.18 &   3.24 &  $-$0.72 &  \nodata &  188 &  170 &  165   \\ 
  FeI  &   5406.78 &   4.37 &  $-$1.62 &   61 &   64 &   58 &   54   \\ 
  FeI  &   5410.92 &   4.47 &   0.40 &  147 &  153 &  134 &  134   \\ 
  FeI  &   5415.21 &   4.39 &   0.64 &  169 &  165 &  149 &  146   \\ 
  FeI  &   5417.04 &   4.41 &  $-$1.58 &   49 &   48 &   56 &   46   \\ 
  FeI  &   5424.08 &   4.32 &   0.51 &  186 &  178 &  169 &  160   \\ 
  FeI  &   5441.33 &   4.10 &  $-$1.63 &   49 &   44 &   51 &   41   \\ 
  FeI  &   5445.05 &   4.39 &  $-$0.03 &  138 &  126 &  120 &  116   \\ 
  FeI  &   5466.39 &   4.37 &  $-$0.62 &  112 &  108 &   97 &   93   \\ 
  FeI  &   5470.09 &   4.44 &  $-$1.71 &   43 &   32 &   29 &   27   \\ 
  FeI  &   5473.90 &   4.15 &  $-$0.69 &  104 &   98 &   97 &   96   \\ 
  FeI  &   5487.14 &   4.41 &  $-$1.43 &  \nodata &   78 &   64 &   55   \\ 
  FeI  &   5493.50 &   4.10 &  $-$1.68 &  \nodata &  \nodata &   82 &   68   \\ 
  FeI  &   5494.46 &   4.07 &  $-$1.99 &   72 &   50 &   48 &   41   \\ 
  FeI  &   5522.45 &   4.21 &  $-$1.45 &   68 &   70 &   59 &   57   \\ 
  FeI  &   5525.55 &   4.23 &  $-$1.08 &   77 &   87 &   71 &   78   \\ 
  FeI  &   5536.58 &   2.83 &  $-$3.71 &  \nodata &  \nodata &   34 &   21   \\ 
  FeI  &   5554.88 &   4.55 &  $-$0.35 &  104 &  109 &  102 &  101   \\ 
  FeI  &   5560.21 &   4.43 &  $-$1.10 &   60 &   64 &   53 &   53   \\ 
  FeI  &   5567.39 &   2.61 &  $-$2.67 &  145 &  143 &  129 &  119   \\ 
  FeI  &   5569.62 &   3.42 &  $-$0.49 &  180 &  177 &  160 &  159   \\ 
  FeI  &   5576.09 &   3.43 &  $-$0.92 &  162 &  157 &  144 &  135   \\ 
  FeI  &   5586.76 &   3.37 &  $-$0.14 &  \nodata &  \nodata &  \nodata &  192   \\  
  FeI  &   5618.63 &   4.21 &  $-$1.63 &   78 &   74 &   68 &   66   \\ 
  FeI  &   5619.59 &   4.39 &  $-$1.53 &  \nodata &  \nodata &   50 &   36   \\ 
  FeI  &   5624.04 &   4.26 &  $-$1.22 &  \nodata &  \nodata &   69 &  \nodata   \\ 
  FeI  &   5641.44 &   4.26 &  $-$1.08 &   95 &  105 &   90 &   87   \\ 
  FeI  &   5650.02 &   5.10 &  $-$0.82 &   38 &   42 &   27 &   36   \\ 
  FeI  &   5650.70 &   5.08 &  $-$0.96 &   31 &   33 &   31 &   29   \\ 
  FeI  &   5652.32 &   4.26 &  $-$1.85 &   41 &   37 &   39 &   33   \\ 
  FeI  &   5653.89 &   4.39 &  $-$1.54 &   58 &   60 &   49 &   48   \\ 
  FeI  &   5661.35 &   4.28 &  $-$1.76 &   58 &   56 &   36 &   37   \\ 
  FeI  &   5662.52 &   4.18 &  $-$0.57 &  140 &  142 &  117 &  112   \\ 
  FeI  &   5679.02 &   4.65 &  $-$0.82 &   71 &   73 &   64 &   66   \\ 
  FeI  &   5680.24 &   4.19 &  $-$2.48 &   34 &   57 &   23 &   19   \\ 
  FeI  &   5698.02 &   3.64 &  $-$2.58 &   59 &   48 &   42 &   37   \\ 
  FeI  &   5701.54 &   2.56 &  $-$2.14 &  172 &  159 &  141 &  130   \\ 
  FeI  &   5705.47 &   4.30 &  $-$1.36 &   58 &   63 &   57 &   50   \\ 
  FeI  &   5731.76 &   4.26 &  $-$1.20 &   85 &   82 &   73 &   68   \\ 
  FeI  &   5741.85 &   4.26 &  $-$1.85 &   63 &   57 &   49 &   42   \\ 
  FeI  &   5752.04 &   4.55 &  $-$0.94 &   76 &   77 &   77 &   64   \\ 
  FeI  &   5753.12 &   4.26 &  $-$0.69 &  113 &  115 &  102 &   97   \\ 
  FeI  &   5760.35 &   3.64 &  $-$2.39 &   60 &   73 &   55 &   39   \\ 
  FeI  &   5762.99 &   4.21 &  $-$0.41 &  149 &  146 &  140 &  129   \\ 
  FeI  &   5775.06 &   4.22 &  $-$1.30 &   85 &   80 &   72 &   75   \\ 
  FeI  &   5778.46 &   2.59 &  $-$3.43 &   89 &   85 &   69 &   67   \\ 
  FeI  &   5793.91 &   4.22 &  $-$1.60 &   55 &   51 &   47 &   52   \\ 
  FeI  &   5827.88 &   3.28 &  $-$3.31 &   45 &   36 &   29 &   20   \\ 
  FeI  &   5838.37 &   3.94 &  $-$2.24 &   43 &   41 &   28 &   34   \\ 
  FeI  &   5852.22 &   4.55 &  $-$1.23 &  \nodata &   86 &   70 &   54   \\ 
  FeI  &   5855.09 &   4.61 &  $-$1.48 &   40 &   44 &   33 &   23   \\ 
  FeI  &   5856.08 &   4.29 &  $-$1.33 &   82 &   76 &   56 &   50   \\ 
  FeI  &   5859.60 &   4.55 &  $-$0.55 &   98 &   98 &   91 &   80   \\ 
  FeI  &   5883.81 &   3.96 &  $-$1.26 &  108 &   94 &   92 &   79   \\ 
  FeI  &   5927.79 &   4.65 &  $-$0.99 &   50 &   51 &   47 &   49   \\ 
  FeI  &   5929.67 &   4.55 &  $-$1.31 &   53 &   47 &   47 &   41   \\ 
  FeI  &   5930.17 &   4.65 &  $-$0.14 &   97 &   91 &   90 &   92   \\ 
  FeI  &   5934.65 &   3.93 &  $-$1.07 &  134 &  125 &  107 &  104   \\ 
  FeI  &   5940.99 &   4.18 &  $-$2.05 &   54 &   52 &   48 &   38   \\ 
  FeI  &   5952.72 &   3.98 &  $-$1.34 &  100 &  101 &   91 &   80   \\ 
  FeI  &   5956.69 &   0.86 &  $-$4.50 &  \nodata &  182 &  151 &  124   \\ 
  FeI  &   5976.79 &   3.94 &  $-$1.33 &  106 &  102 &   96 &   91   \\ 
  FeI  &   5983.69 &   4.55 &  $-$0.66 &  106 &   98 &   90 &   91   \\ 
  FeI  &   5984.83 &   4.73 &  $-$0.26 &  \nodata &  \nodata &  109 &   97   \\ 
  FeI  &   6024.05 &   4.55 &   0.03 &  120 &  118 &  119 &  114   \\ 
  FeI  &   6027.05 &   4.07 &  $-$1.09 &   95 &   97 &   85 &   80   \\ 
  FeI  &   6055.99 &   4.73 &  $-$0.37 &   86 &   86 &   81 &   80   \\ 
  FeI  &   6065.48 &   2.61 &  $-$1.41 &  \nodata &  \nodata &  179 &  166   \\ 
  FeI  &   6078.50 &   4.79 &  $-$0.33 &   90 &   83 &   86 &   78   \\ 
  FeI  &   6079.00 &   4.65 &  $-$1.02 &   65 &   59 &   57 &   48   \\ 
  FeI  &   6089.57 &   5.02 &  $-$0.90 &  \nodata &   63 &   58 &   55   \\ 
  FeI  &   6093.67 &   4.65 &  $-$1.40 &   47 &   45 &   45 &  \nodata   \\ 
  FeI  &   6094.37 &   4.65 &  $-$1.84 &   28 &   27 &   35 &   19   \\ 
  FeI  &   6096.66 &   3.98 &  $-$1.83 &   68 &   62 &   63 &   57   \\ 
  FeI  &   6137.69 &   2.59 &  $-$1.35 &  \nodata &  \nodata &  \nodata &  195   \\ 
  FeI  &   6151.62 &   2.18 &  $-$3.37 &  142 &  137 &  114 &  101   \\ 
  FeI  &   6157.73 &   4.07 &  $-$1.16 &  111 &  113 &  102 &   93   \\ 
  FeI  &   6165.36 &   4.14 &  $-$1.47 &   81 &   69 &   64 &   58   \\ 
  FeI  &   6173.34 &   2.22 &  $-$2.88 &  170 &  149 &  131 &  125   \\ 
  FeI  &   6180.20 &   2.73 &  $-$2.65 &  134 &  126 &  104 &   99   \\ 
  FeI  &   6187.99 &   3.94 &  $-$1.62 &   84 &   89 &   72 &   68   \\ 
  FeI  &   6200.31 &   2.61 &  $-$2.37 &  144 &  148 &  129 &  118   \\ 
  FeI  &   6240.65 &   2.22 &  $-$3.17 &  142 &  132 &  108 &  100   \\ 
  FeI  &   6246.32 &   3.60 &  $-$0.88 &  150 &  145 &  136 &  136   \\ 
  FeI  &   6252.55 &   2.40 &  $-$1.77 &  \nodata &  \nodata &  194 &  176   \\ 
  FeI  &   6254.26 &   2.28 &  $-$2.43 &  192 &  178 &  169 &  155   \\ 
  FeI  &   6265.13 &   2.18 &  $-$2.54 &  195 &  190 &  160 &  145   \\ 
  FeI  &   6271.28 &   3.33 &  $-$2.70 &   70 &   71 &   58 &   44   \\ 
  FeI  &   6290.97 &   4.73 &  $-$0.73 &   77 &   81 &   77 &   71   \\ 
  FeI  &   6297.79 &   2.22 &  $-$2.64 &  175 &  167 &  130 &  137   \\ 
  FeI  &   6301.51 &   3.65 &  $-$0.72 &  154 &  155 &  139 &  144   \\ 
  FeI  &   6302.50 &   3.69 &  $-$1.11 &  146 &  142 &  131 &  105   \\ 
  FeI  &   6311.50 &   2.83 &  $-$3.14 &  101 &   99 &   72 &   58   \\ 
  FeI  &   6380.75 &   4.19 &  $-$1.38 &   86 &   87 &   78 &   73   \\ 
  FeI  &   6392.54 &   2.28 &  $-$3.99 &   93 &   85 &   64 &   58   \\ 
  FeI  &   6393.60 &   2.43 &  $-$1.58 &  \nodata &  \nodata &  \nodata &  186   \\
  FeI  &   6408.03 &   3.69 &  $-$1.02 &  141 &  132 &  128 &  118   \\ 
  FeI  &   6411.65 &   3.65 &  $-$0.72 &  161 &  154 &  142 &  140   \\ 
  FeI  &   6421.35 &   2.28 &  $-$2.01 &  \nodata &  \nodata &  188 &  175   \\ 
  FeI  &   6469.21 &   4.83 &  $-$0.73 &  100 &   98 &   81 &   76   \\ 
  FeI  &   6475.63 &   2.56 &  $-$2.94 &  135 &  125 &  112 &   89   \\ 
  FeI  &   6481.87 &   2.28 &  $-$3.01 &  156 &  139 &  128 &  113   \\ 
  FeI  &   6483.94 &   1.48 &  $-$5.34 &   61 &   46 &   41 &   23   \\ 
  FeI  &   6495.74 &   4.83 &  $-$0.84 &   61 &   64 &   56 &   41   \\ 
  FeI  &   6498.94 &   0.96 &  $-$4.69 &  196 &  188 &  153 &  123   \\ 
  FeI  &   6533.93 &   4.56 &  $-$1.36 &   53 &   62 &   56 &   49   \\ 
  FeI  &   6581.21 &   1.48 &  $-$4.68 &  148 &  100 &   83 &   71   \\ 
  FeI  &   6592.91 &   2.73 &  $-$1.47 &  198 &  171 &  160 &  158   \\ 
  FeI  &   6593.87 &   2.43 &  $-$2.37 &  178 &  142 &  134 &  128   \\ 
  FeI  &   6608.02 &   2.28 &  $-$3.93 &  103 &   88 &   66 &   51   \\ 
  FeI  &   6609.11 &   2.56 &  $-$2.66 &  155 &  148 &  118 &  115   \\ 
  FeI  &   6625.02 &   1.01 &  $-$5.37 &  \nodata &  177 &  118 &   83   \\ 
  FeI  &   6627.54 &   4.79 &  $-$1.58 &   47 &   36 &   46 &  \nodata   \\ 
  FeI  &   6633.75 &   4.79 &  $-$0.80 &  \nodata &   87 &   86 &   82   \\ 
  FeI  &   6646.93 &   2.61 &  $-$3.96 &   65 &   65 &   50 &   41   \\ 
  FeI  &   6648.12 &   1.01 &  $-$5.92 &  103 &   91 &   69 &   48   \\ 
  FeI  &   6713.77 &   4.79 &  $-$1.50 &   25 &   35 &   30 &   20   \\ 
  FeI  &   6715.38 &   4.61 &  $-$1.54 &   44 &   44 &   39 &   29   \\ 
  FeI  &   6716.22 &   4.58 &  $-$1.85 &   31 &   35 &   28 &   18   \\ 
  FeI  &   6725.35 &   4.19 &  $-$2.25 &   42 &   39 &   34 &   28   \\ 
  FeI  &   6726.67 &   4.61 &  $-$1.07 &   58 &   62 &   58 &   53   \\ 
  FeI  &   6733.15 &   4.64 &  $-$1.48 &   33 &   40 &   37 &   36   \\ 
  FeI  &   6739.52 &   1.56 &  $-$4.79 &  103 &   96 &   74 &   54   \\ 
  FeI  &   6750.15 &   2.42 &  $-$2.58 &  172 &  158 &  143 &  133   \\ 
  FeI  &   6752.71 &   4.64 &  $-$1.20 &  \nodata &  \nodata &   72 &   52   \\ 
  FeI  &   6783.71 &   2.59 &  $-$3.92 &   71 &   70 &   53 &   27   \\ 
  FeI  &   6786.86 &   4.19 &  $-$1.97 &   37 &   65 &   48 &   32   \\ 
  FeI  &   6837.02 &   4.59 &  $-$1.69 &   28 &   26 &   24 &   25   \\ 
  FeI  &   6839.83 &   2.56 &  $-$3.35 &  105 &  100 &   83 &   74   \\ 
  FeI  &   6842.68 &   4.64 &  $-$1.22 &   58 &   53 &   53 &   45   \\ 
  FeI  &   6843.65 &   4.55 &  $-$0.83 &   79 &   68 &   70 &   71   \\ 
  FeI  &   6851.63 &   1.61 &  $-$5.28 &   70 &   71 &   41 &   27   \\ 
  FeI  &   6855.18 &   4.56 &  $-$0.74 &  100 &   96 &   98 &   88   \\ 
  FeI  &   6855.71 &   4.61 &  $-$1.78 &  \nodata &  \nodata &   36 &   35   \\ 
  FeI  &   6858.15 &   4.61 &  $-$0.93 &   68 &   67 &   68 &   63   \\ 
  FeI  &   6861.95 &   2.42 &  $-$3.85 &   93 &   80 &   71 &   56   \\ 
  FeI  &   6862.49 &   4.56 &  $-$1.47 &  \nodata &  \nodata &   41 &   37   \\ 
  FeI  &   6971.93 &   3.02 &  $-$3.34 &   59 &   51 &   39 &   32   \\ 
  FeI  &   6978.85 &   2.48 &  $-$2.45 &  \nodata &  \nodata &  150 &  138   \\ 
  FeI  &   6988.52 &   2.40 &  $-$3.56 &  116 &  110 &   96 &   81   \\ 
  FeI  &   6999.88 &   4.10 &  $-$1.46 &   82 &   89 &   70 &   74   \\ 
  FeI  &   7000.62 &   4.14 &  $-$2.39 &   37 &   35 &   39 &   27   \\ 
  FeI  &   7014.98 &   4.19 &  $-$4.20 &   69 &   50 &   57 &  \nodata   \\ 
  FeI  &   7022.95 &   4.19 &  $-$1.15 &   98 &   86 &   85 &   87   \\ 
  FeI  &   7038.22 &   4.22 &  $-$1.20 &   97 &  102 &   83 &   80   \\ 
  FeII &   4923.93 &   3.23 &  $-$1.32 &  168 &  173 &  177 &  \nodata   \\ 
  FeII &   5197.58 &   3.22 &  $-$2.23 &  \nodata &  \nodata &   88 &  133   \\ 
  FeII &   5234.63 &   3.22 &  $-$2.22 &   89 &   83 &   85 &  102   \\ 
  FeII &   5414.08 &   3.22 &  $-$3.62 &   21 &   38 &   44 &   43   \\ 
  FeII &   5534.85 &   3.25 &  $-$2.64 &  \nodata &  \nodata &   98 &   89   \\ 
  FeII &   6084.11 &   3.20 &  $-$3.80 &   19 &   38 &   40 &   30   \\ 
  FeII &   6149.26 &   3.89 &  $-$2.69 &   38 &   34 &   36 &   51   \\ 
  FeII &   6247.56 &   3.89 &  $-$2.36 &   43 &   42 &   58 &   69   \\ 
  FeII &   6369.46 &   2.89 &  $-$4.20 &   23 &   27 &   26 &   34   \\ 
  FeII &   6416.92 &   3.89 &  $-$2.69 &   36 &   42 &   46 &   47   \\ 
  CoI  &   6516.08 &   1.71 &  $-$3.45 &   61 &   43 &   74 &  \nodata   \\ 
  CoI  &   5530.79 &   1.71 &  $-$2.06 &   92 &   81 &   57 &   51   \\ 
  CoI  &   5647.23 &   2.28 &  $-$1.56 &   58 &   51 &   35 &   25   \\ 
  CoI  &   6189.00 &   1.71 &  $-$2.45 &   89 &   91 &   55 &   42   \\ 
  CoI  &   6632.45 &   2.28 &  $-$2.00 &   58 &   46 &   37 &   27   \\ 
  NiI  &   5578.72 &   1.68 &  $-$2.64 &  148 &  131 &  116 &  100   \\ 
  NiI  &   5587.86 &   1.93 &  $-$2.14 &  155 &  134 &  108 &  101   \\ 
  NiI  &   5589.36 &   3.90 &  $-$1.14 &   38 &   32 &   30 &   22   \\ 
  NiI  &   5593.74 &   3.90 &  $-$0.84 &   47 &   48 &   43 &   42   \\ 
  NiI  &   5682.20 &   4.10 &  $-$0.47 &   61 &   64 &   44 &   49   \\ 
  NiI  &   5748.35 &   1.68 &  $-$3.26 &  112 &  106 &   84 &   67   \\ 
  NiI  &   5796.09 &   1.95 &  $-$3.69 &   51 &   43 &   36 &   27   \\ 
  NiI  &   5805.22 &   1.68 &  $-$0.64 &  \nodata &  \nodata &   37 &  \nodata   \\ 
  NiI  &   5846.99 &   1.68 &  $-$3.21 &  105 &   98 &   87 &   56   \\ 
  NiI  &   6053.69 &   4.23 &  $-$1.07 &   28 &   23 &   26 &   14   \\ 
  NiI  &   6128.97 &   1.68 &  $-$3.33 &   99 &   93 &   72 &   62   \\ 
  NiI  &   6175.37 &   4.09 &  $-$0.54 &   57 &   54 &   46 &   47   \\ 
  NiI  &   6176.81 &   4.09 &  $-$0.53 &   72 &   68 &   64 &   66   \\ 
  NiI  &   6177.24 &   1.83 &  $-$3.51 &   76 &   63 &   50 &   35   \\ 
  NiI  &   6314.66 &   3.54 &  $-$1.77 &  \nodata &  136 &  115 &  \nodata   \\ 
  NiI  &   6370.35 &   3.54 &  $-$1.94 &   22 &   21 &   19 &   16   \\ 
  NiI  &   6378.25 &   4.15 &  $-$0.90 &   39 &   36 &   33 &   30   \\ 
  NiI  &   6482.80 &   1.93 &  $-$2.63 &  114 &  101 &   90 &   82   \\ 
  NiI  &   6635.12 &   4.42 &  $-$0.83 &   26 &   18 &   22 &   18   \\ 
  NiI  &   6643.63 &   1.68 &  $-$2.30 &  192 &  179 &  153 &  150   \\ 
  NiI  &   6767.77 &   1.83 &  $-$2.17 &  162 &  152 &  134 &  127   \\ 
  NiI  &   6772.31 &   3.66 &  $-$0.99 &   67 &   70 &   57 &   56   \\ 
  NiI  &   6842.04 &   3.66 &  $-$1.47 &   53 &   44 &   38 &   33   \\ 
  CuI  &   5105.54 &   1.39 &  $-$1.50 &  177 &  174 &  149 &  120   \\ 
  CuI  &   5782.12 &   1.64 &  $-$1.78 &  142 &  140 &  111 &  101   \\ 
  ZnI  &   4722.16 &   4.03 &  $-$0.39 &  \nodata &   39 &   56 &   63   \\ 
  ZnI  &   4810.54 &   4.08 &  $-$0.17 &   57 &   56 &   53 &   67   \\ 
  YII  &   4883.69 &   1.08 &   0.07 &  153 &  115 &  103 &   98   \\ 
  YII  &   5087.43 &   1.08 &  $-$0.17 &   84 &   71 &   72 &   78   \\ 
  YII  &   5200.42 &   0.99 &  $-$0.57 &  \nodata &  \nodata &  \nodata &   81   \\  
  ZrI  &   6127.44 &   0.15 &  $-$1.06 &  102 &   75 &   32 &   16   \\ 
  ZrI  &   6134.55 &   0.00 &  $-$1.28 &  105 &   73 &   28 &   13   \\ 
  ZrI  &   6143.20 &   0.07 &  $-$1.10 &  104 &   78 &   37 &   16   \\ 
  BaII &   5853.70 &   0.60 &  $-$1.01 &  167 &  160 &  140 &  132   \\ 
  BaII &   6141.70 &   0.70 &  $-$0.07 &  242 &  234 &  199 &  193   \\ 
  BaII &   6496.90 &   0.60 &  $-$0.38 &  238 &  224 &  198 &  193   \\ 
  LaII &   6390.48 &   0.32 &  $-$1.41 &   75 &   65 &   48 &   30   \\ 
  LaII &   6774.26 &   0.13 &  $-$1.72 &   83 &   66 &   53 &   30   \\ 
  NdII &   5130.59 &   1.30 &   0.45 &  \nodata &   86 &   57 &   63   \\ 
  NdII &   5319.81 &   0.55 &  $-$0.14 &  107 &   97 &   85 &   82   \\ 
  EuII &   6645.11 &   1.38 &   0.12 &   72 &   62 &   46 &   56   \\ 
\enddata
\end{deluxetable}

%
%
\clearpage                                                                      
\begin{deluxetable}{l rrc rrc rrc rrc}                             
\tablenum{4}                                                                    
\tablewidth{0pt}                                                                
\tablecaption{Derived Abundances for Pal 12 
\label{table_abund}}                      
\tablehead{ 
\colhead{Star} & \colhead{S1} & \colhead{} & \colhead{} &
\colhead{1118} & \colhead{} & \colhead{} &
\colhead{1128} & \colhead{} & \colhead{} &
\colhead{1305} & \colhead{} & \colhead{} \\                                                                    
\colhead{} & 
\colhead{[X/Fe]} & \colhead{$\sigma$\tablenotemark{a}} & \colhead{No.} &
\colhead{[X/Fe]} & \colhead{$\sigma$\tablenotemark{a}} & \colhead{No.} &
\colhead{[X/Fe]} & \colhead{$\sigma$\tablenotemark{a}} & \colhead{No.} &
\colhead{[X/Fe]} & \colhead{$\sigma$\tablenotemark{a}} & \colhead{No.} \\
\colhead{Ion} & 
\colhead{(dex)} & \colhead{(dex)} & \colhead{Lines} &
\colhead{(dex)} & \colhead{(dex)} & \colhead{Lines} &
\colhead{(dex)} & \colhead{(dex)} & \colhead{LInes} &
\colhead{(dex)} & \colhead{(dex)} & \colhead{Lines} 
}
\startdata
  OI   &  0.04 &   0.04 &    2 &   0.21 &   0.19 &    2 &  $-$0.06 &   0.21 &    2 &   0.07 &   0.13 &  2 \\
  NaI  &  $-$0.47 &   0.18 &    4 &  $-$0.52 &   0.16 &    4 &  $-$0.55 &   0.20 &    4 &  $-$0.49 &   0.16 &    4  \\ 
  MgI  &  $-$0.14 &  \nodata &    1 &   0.07 &   0.11 &    2 &   0.09 &   0.30 &    2 &   0.09 &   0.18 &    3  \\ 
  SiI  &   0.14 &   0.16 &   12 &   0.03 &   0.16 &   13 &   0.11 &   0.18 &   13 &   0.10 &   0.14 &   13  \\ 
  CaI  &  $-$0.17 &   0.24 &   13 &  $-$0.25 &   0.22 &   14 &  $-$0.18 &  0.19 &   14 &  $-$0.16 &   0.15 &   14  \\ 
  ScII &  $-$0.16 &   0.17 &    6 &  $-$0.12 &   0.16 &    6 &  $-$0.08 &   0.16 &    7 &  $-$0.07 &   0.17 &    7  \\ 
  TiI  &  $-$0.10 &   0.21 &   26 &  $-$0.17 &   0.16 &   28 &  $-$0.12 &   0.25 &   33 &  $-$0.12 &   0.14 &   30  \\ 
  TiII &   0.02 &   0.25 &    4 &  $-$0.03 &   0.13 &    5 &   0.08 &   0.18 &    6 &   0.08 &   0.10 &    5  \\ 
  VI   &  $-$0.31 &   0.19 &    8 &  $-$0.41 &   0.20 &    9 &  $-$0.35 &   0.15 &   10 &  $-$0.38 &   0.14 &   10  \\ 
  CrI  &   0.14 &   0.21 &    6 &   0.02 &   0.23 &    6 &  $-$0.03 &   0.20 &   10 &   0.07 &   0.21 &   10  \\ 
  MnI  &  $-$0.25 &   0.13 &    2 &  $-$0.28 &   0.09 &    2 &  $-$0.32 &   0.11 &    4 &  $-$0.28 &   0.20 &    5  \\ 
  FeI\tablenotemark{b}  &   
    6.76 &   0.18 &  123 &   6.72 &   0.23 &  131 &   6.70 &   0.19 &  146 &   6.72 &   0.19 &  146  \\ 
  FeII &   0.10 &   0.23 &    9 &   0.09 &   0.19 &    9 &   0.12 &   0.23 &   11 &   0.12 &   0.20 &    9  \\ 
  CoI  &  $-$0.35 &   0.21 &    4 &  $-$0.32 &   0.28 &    4 &  $-$0.29 &   0.21 &    4 &  $-$0.25 &   0.19 &    4  \\ 
  NiI  &  $-$0.11 &   0.17 &   21 &  $-$0.21 &   0.17 &   22 &  $-$0.20 &   0.19 &   23 &  $-$0.20 &   0.18 &   21  \\ 
  CuI  &  $-$0.55 &   0.18 &    2 &   0.14 &   0.16 &    2 &  $-$1.01 &   0.27 &    2 &  $-$0.69 &   0.07 &    2  \\ 
  ZnI  &  $-$0.45 &  \nodata &    1 &  $-$0.64 &   0.16 &    2 &  $-$0.51 &   0.18 &    2 &  $-$0.38 &   0.06 &    2  \\ 
  YII  &  $-$0.16 &   0.93 &    2 &  $-$0.60 &   0.52 &    2 &  $-$0.49 &   0.36 &    2 &  $-$0.35 &   0.17 &    3  \\ 
  ZrI  &  $-$0.16 &   0.07 &    3 &  $-$0.28 &   0.06 &    3 &  $-$0.22 &   0.05 &    3 &  $-$0.16 &   0.06 &    3  \\ 
  BaII &   0.28 &   0.05 &    3 &   0.27 &   0.05 &    3 &   0.25 &   0.06 &    3 &   0.27 &   0.08 &    3  \\ 
  LaII &   0.25 &   0.06 &    2 &   0.22 &   0.01 &    2 &   0.31 &   0.08 &    2 &   0.09 &   0.02 &    2  \\ 
  NdII &   0.27 &  \nodata &    1 &   0.38 &   0.26 &  2 &   0.28 &   0.06 &    2 &   0.40 &   0.05 &   2  \\ 
  EuII &   0.62 &  \nodata &    1 &   0.58 &  \nodata &    1 &   0.55 &  \nodata &    1 &   0.69 &  \nodata &    1  \\ 
\enddata
\tablenotetext{a}{This is the 1$\sigma$ rms deviation of the set of 
abundances derived from
each of the observed absorption lines about the mean abundance.}
\tablenotetext{b}{For Fe~I only, we give [Fe/H].}
\end{deluxetable}

%
%
\clearpage                                                                      
\begin{deluxetable}{l rrc rrc}                             
\tablenum{5}                                                                    
\tablewidth{0pt}                                                                
\tablecaption{Mean Abundances and Abundance Spreads for Four Stars in Pal 12 
\label{table_abundsig}}                      
\tablehead{ 
\colhead{Species} & \colhead{Mean Abund.} &
\colhead{$\sigma$ Around Mean} & \colhead{$\sigma$(Obs)} &
\colhead{Spread Ratio\tablenotemark{a}} & \colhead{No. of Stars\tablenotemark{b}} \\
\colhead{} & \colhead{[X/Fe] (dex)} &\colhead{(dex)} &
\colhead{(dex)} & \colhead{(dex)} \\
}
\startdata
OI   &    0.07 &    0.11 &    0.10 &    1.10 &  4 \\
NaI  &   $-$0.51 &    0.04 &    0.09 &    0.42 &  4 \\
MgI  &    0.08 &    0.01 &    0.10 &    0.11 &  3 \\
SiI  &    0.10 &    0.05 &    0.05\tablenotemark{c} &  0.90 &  4 \\
CaI  &   $-$0.19 &    0.04 &    0.05 &    0.76 &  4 \\
ScII &   $-$0.11 &    0.04 &    0.07 &    0.62 &  4 \\
TiI  &   $-$0.13 &    0.03 &    0.05\tablenotemark{c} & 0.58 &  4 \\
TiII &    0.04 &    0.05 &    0.07 &    0.72 &  4 \\
VI   &   $-$0.36 &    0.04 &    0.06 &    0.78 &  4 \\
CrI  &    0.05 &    0.07 &    0.09 &    0.83 &  4 \\
MnI  &   $-$0.28 &    0.03 &    0.09 &    0.30 &  4 \\
FeI\tablenotemark{e}  &    6.72 &    0.02 &    0.05\tablenotemark{c} &  0.44 &  4 \\
FeII &    0.11 &    0.02 &    0.07 &    0.23 &  4 \\
CoI  &   $-$0.30 &    0.04 &    0.11 &    0.39 &  4 \\
NiI  &   $-$0.18 &    0.04 &    0.05\tablenotemark{c} & 0.90 &  4 \\
CuI  &   $-$0.52 &    0.49 &    0.12 &    4.05\tablenotemark{d} &  4 \\
ZnI  &   $-$0.51 &    0.13 &    0.07 &    1.86 &  3 \\
YII  &   $-$0.48\tablenotemark{f} &    0.12 &    0.19 &    0.67 &  3 \\
ZrI   &   $-$0.20 &    0.06 &    0.05\tablenotemark{c} & 1.12 &  4 \\
BaII &    0.27 &    0.01 &    0.05\tablenotemark{c} &    0.24 &  4 \\
LaII &    0.22 &    0.09 &    0.05\tablenotemark{c} &  1.80 &  4 \\
NdII &    0.35 &    0.07 &    0.07 &    1.06 &  3 \\
EuII &    0.61 &    0.06 & \nodata & \nodata     &  4 \\
\enddata
\tablenotetext{a}{This is the ratio of $\sigma$ about the mean abundance
for Pal 12 of the sample of four stars to $\sigma$(Obs).} 
\tablenotetext{b}{For some species, Star S1, the coolest star 
in our Pal 12 sample, had significantly
fewer usable lines, as the spectrum was more crowded and the lines
became too strong to use.}
\tablenotetext{c}{$\sigma$(obs) is very low, in most cases due
to the large number of lines used.  It has been increased to 0.05 dex.}
\tablenotetext{d}{Cu~I has very large HFS corrections, between
0.5 and 1.0 dex.}
\tablenotetext{e}{For Fe~I only, we give [Fe/H].}
\tablenotetext{f}{The coolest star is excluded; see \S\ref{sec_individual}.}
\end{deluxetable}

\end{document}